\documentclass{collective-intelligence}

\usepackage[normalem]{ulem}
\usepackage{graphicx}
\usepackage{color}
\usepackage{harvard}

\title{CROWDSOURCING GAZE DATA COLLECTION}
\numberofauthors{4} 
\author{
\alignauthor Dmitry Rudoy\\
       \affaddr{Technion}\\
       \email{dmitryr@tx.technion.ac.il}
\alignauthor Dan B Goldman\\
       \affaddr{Adobe Systems}\\
       \email{dgoldman@adobe.com}
\and  
\alignauthor Eli Shechtman\\
       \affaddr{Adobe Systems}\\
       \email{elishe@adobe.com}
\alignauthor Lihi Zelnik-Manor\\
       \affaddr{Technion}\\
       \email{lihi@ee.technion.ac.il}
}

\begin{document}
\normalem

\teaser{
    \centering
    \begin{tabular}{ccc}
        \includegraphics[height=3cm]{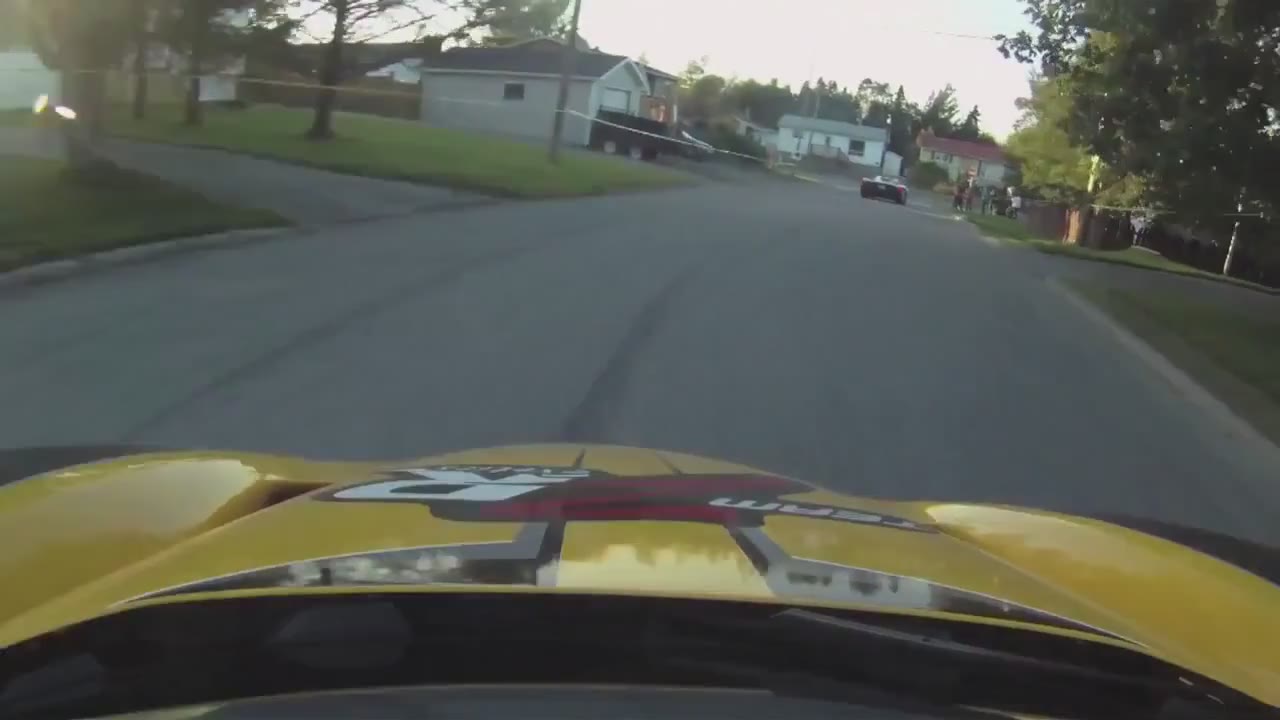} &
        \includegraphics[height=3cm]{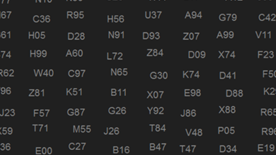} &
        \includegraphics[height=3cm]{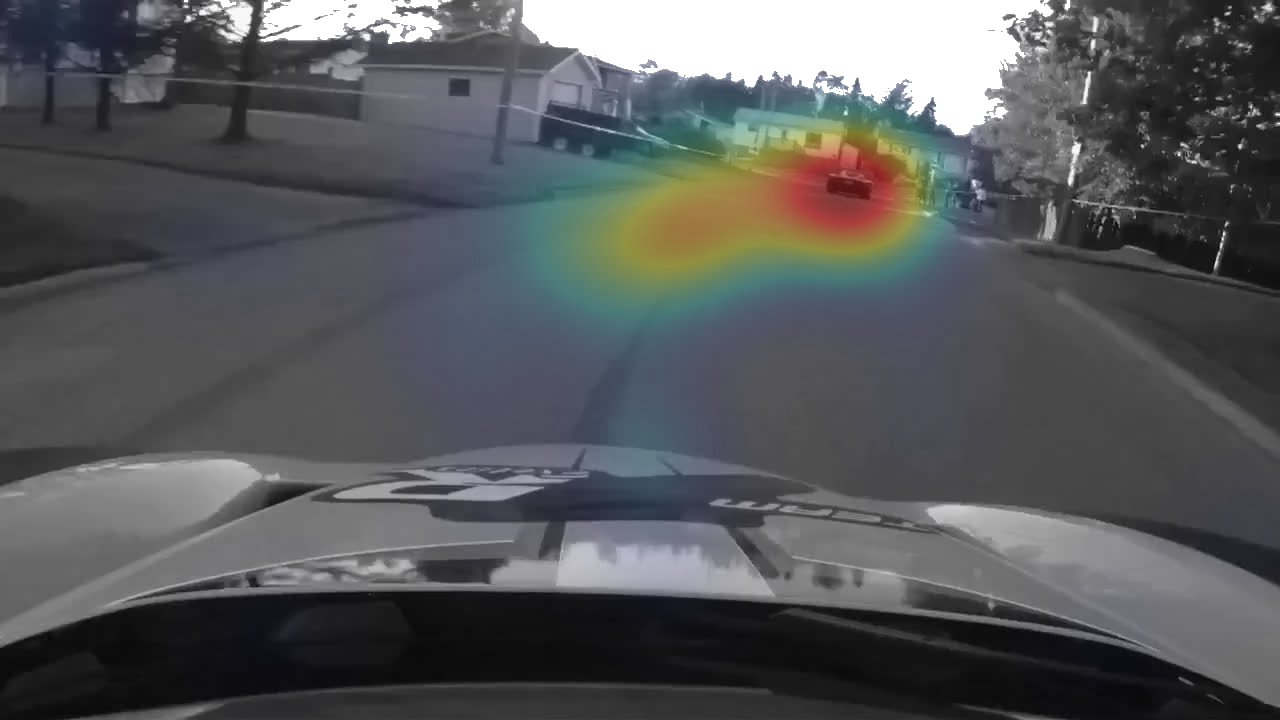}
    \end{tabular}
    \caption{\label{fig:teaser}
    Our system enables collection of gaze direction data from any video information (left), from any number of participants. It employs a character chart (center) to validate self-reported gaze location and aggregates data from multiple participants into a probability density function, visualized on the right as a heat map.
    }
}

\maketitle
\begin{abstract}
Knowing where people look is a useful tool in many various image and video
applications. However, traditional gaze tracking hardware is expensive and
requires local study participants, so acquiring gaze location data from a
large number of participants is very problematic. In this work we propose a
crowdsourced method for acquisition of gaze direction data from a virtually
unlimited number of participants, using a robust self-reporting mechanism (see Figure~\ref{fig:teaser}). Our
system collects temporally sparse but spatially dense points-of-attention in
any visual information. We apply our approach to an existing video data set
and demonstrate that we obtain results similar to traditional gaze tracking.
We also explore the parameter ranges of our method, and collect gaze tracking
data for a large set of YouTube videos.
\end{abstract}


\section{Introduction}
\label{intro}

The visual design of images, videos and human-computer interfaces are often
explicitly constructed to draw an observer's attention to specific regions of
a screen at desired moments in time. By manipulating brightness, contrast,
color, motion, texture, focus, and structure, the designer or cinematographer
may attempt to induce a particular gaze direction in the viewer. And, because
the fovea - the portion of the retina corresponding to the gaze direction -
contains a much higher density of receptors than the rest of the retina, this
gaze direction also indicates the part of the image from which the viewer
receives the most visual information. Thus, designers benefit from knowing
where the end consumers of their designs are actually looking.

It is sometimes possible to record the gaze direction of human participants
using gaze tracking hardware. Although the idea of tracking the human eye is
very old, accurate devices have appeared only during recent decades. Advanced
desktop and head-mounted gaze tracking devices~\cite{srresearch}, or even mobile
glasses~\cite{smi} are now commercially available. Those devices are
capable of producing a dense stream of high accuracy gaze tracking data given
any visual stimulus.

However, gaze tracking hardware is not suitable for all situations. First, in
spite of improved affordability, the hardware itself remains too expensive for
anyone other than scientists and researchers to use. Second, in the lab
setting, only a limited number of individuals can participate in any given
experiment, as they must be physically present at the location of the
hardware, and can only use it one at a time. Third, for video or interactive
materials the experiment may take a relatively long time, even if gaze
direction is only of interest at sparse points in time. Fourth, the limited
deployment of such devices makes it hard to collect statistically significant
amounts of gaze data from widely varying demographics.

In this work we propose a method to acquire gaze location
\footnote{We use the term ``gaze location" to refer to a pixel on the screen where the
viewer focuses his attention. In contrast, note that gaze tracking hardware
typically acquires ``gaze direction," which can be transformed to gaze location
via calibration of the observer's position relative to the screen.}
reliably, without
the need for custom gaze tracking hardware. We focus specifically on video
stimuli, because the expense of gaze tracking hardware has limited the
acquisition of large-scale gaze tracked video databases. Our method acquires
gaze location during dynamic stimuli using ubiquitous hardware, available
throughout the world, and software delivered over the Internet. Our method
relies on self-reporting of gaze direction, using a variety of techniques to
ensure reliability and robustness of the self-reported data. This allows us
both to lower the overall cost of the experiments and to gather data from a
virtually unlimited number of participants.


\section{Related Work}
\label{related}

Gaze tracking has been widely studied in psychology and perception over the last decades. Recent psychological studies show that the spatial distribution of gaze is different in static and dynamic scenes~\cite{goldstein2007people}. There is evidence that the distribution of human gaze directions is more tightly clustered while watching movies compared to images~\cite{mital2010clustering}. Furthermore, the gaze direction is strongly affected by the motion in the scene. Smith and Henderson~\citeyear{smith2008edit} use gaze tracking to validate the ``edit blindness" phenomena -- the fact that humans do not always pay attention to scene cuts.

Another usage of gaze tracking is human-computer interaction.
For instance, Rele and Duchowski~\citeyear{rele2005using} find gaze tracking mechanism useful for validating and improving search results. Specifically, they discuss the importance of different interface elements based on where subjects look.

Computer vision research uses gaze tracking data as well. Judd et al.~\citeyear{judd2009learning} construct a predictive model of where viewers look in static images using gaze tracking. Nataraju et al.~\citeyear{nataraju2009learning} build a similar model for learning attention points in video, also employing gaze tracking data. Both works employ special hardware in a laboratory to obtain the gaze tracking data.


\section{Data Collection Method}
\label{gazetrack}

Before presenting the proposed crowdsourced gaze data collection method we define the goal of the approach. Given a video clip as an input our system records for any arbitrary participant the point on the screen closest to the center of foveation -- the gaze location -- while watching the video. Since this location changes over time we may wish to sample it at any particular video frame.

\subsection{Gaze Location Estimation Method}

\begin{figure*}[htb]
    \centering
    \includegraphics[width=0.8\linewidth]{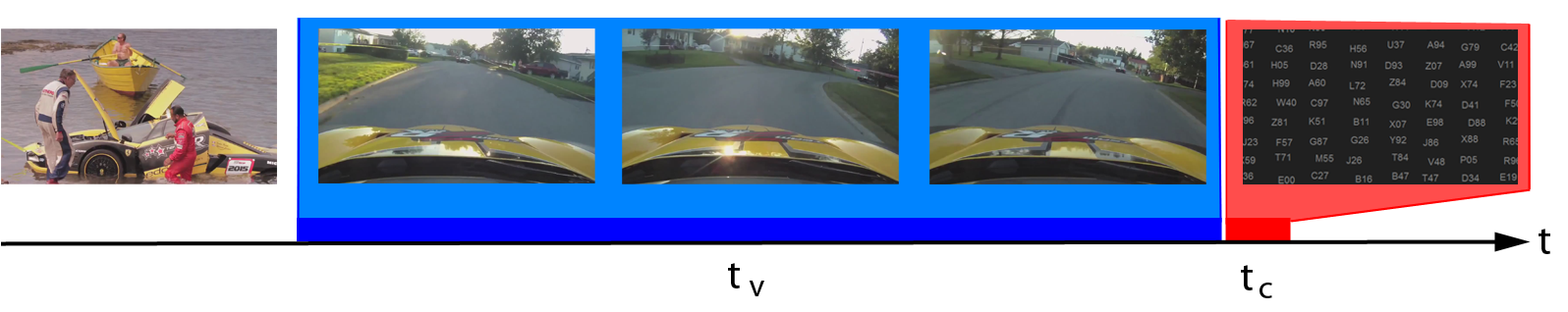}
  \caption{\label{fig:gazetrack}
Scheme of the proposed gaze tracking system. Instead of playing the entire video we show only its short portion $t_v$ prior the frame of interest (blue highlight). After the clip is finished the character chart replaces it for a short time $t_c$ (red highlight) We use $t_v=10sec$ and $t_c=1sec$.
}
\end{figure*}

In contrast to traditional gaze tracking approaches, our system does not aim to collect gaze location for every frame of the given video. Instead, we only record the gaze on a single video frame. To do so we show the user a small portion of the input video, of length $t_v$ seconds.
Immediately after the video ends we replace it with a chart of letter and number symbol combinations that covers the same portion of the screen as the original video. The chart is displayed for a short time of $t_c$ seconds and disappears afterwards. Then the user is asked to answer which letter and number symbols he has seen most clearly. These are entered into a blank text box without any default value, allowing us to screen invalid responses due to inattentive participants. If the combination exists in the chart that was displayed, its location is recorded as the gaze location. The system is sketched in Figure~\ref{fig:gazetrack}.

Since the character chart is used to capture the viewer's gaze location, it should cover the entire video frame uniformly. To build it we place a virtual grid of the same size as the video frame with nodes placed according to the desired density, $D_r$. (The choice of the density is discussed later.) We initially place a triplet of characters at every node of the grid: one letter followed by two digits. The letter and the numbers are chosen randomly, with the letters ``I" and ``O" excluded to prevent confusion with numbers ``1" and ``0".
To break the regularity of the grid we jitter the triplets randomly around the nodes' locations, while ensuring that they do not overlap. Our goal is to capture a wide range of participants at low cost without requiring software installation. Therefore we deployed our gaze tracking software using
Amazon Mechanical Turk~\footnote{http://www.mturk.com}, and implemented the experiment as a Flash application that can play in over 99\% of internet-connected PCs~\footnote{http://www.adobe.com}.

\begin{figure}[htb]
    \centering
	\includegraphics[height=4cm]{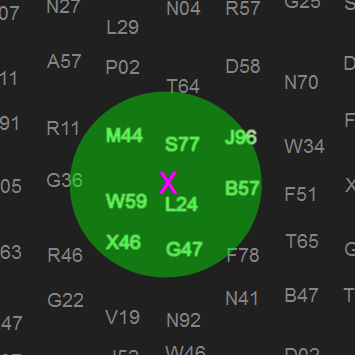}
  \caption{\label{fig:tutorial}
The tutorial for user training. We use a moving colored capital letter ``X" that the user should follow. When the ``X" is replaced with the character chart, the participant should note out its location by entering the closest triplet of letters and numbers. The tutorial is approved if the user's self-reported gaze location is inside the green circle. (The ``X" is not shown at the same time as the chart -- it is overlayed on the chart in this figure for visualization only.)
}
\end{figure}

Since we deploy our system on Mechanical Turk, the participants have highly varying abilities and levels of interest. Thus we provide textual instruction and also require the user to perform a mandatory tutorial at the beginning of every session. The tutorial has the same structure as the gaze tracking mechanism, but instead of the video clip we show the user a moving letter.
The letter follows a random path with constant speed but gradually changing direction over a duration of $t_t = 3$ seconds. After the end of the animation we show the character chart for $t_c$ seconds.
In this case, after the participant indicates the characters they saw closest to their gaze direction, we check if they fall close enough to the last location of the moving letter (see Figure~\ref{fig:tutorial}). If the distance between the indicated location and the true one is smaller than $R_a$ pixels, the tutorial is approved. The user must pass two such tutorial sessions before proceeding to the video experiment. If the total number of attempted tutorials exceeds 10, the participant is rejected.

To add even more randomness to the tutorial we randomize the color of the moving letter using the following colors: red, green, yellow, cyan and magenta. All those colors are high contrast on black background.

%


\section{Experimental Validation}
\label{exp}
In this section we describe our experiment validating the accuracy of the proposed gaze data collection method.
First we describe the overall process of the crowdsourced data collection. Then we present the comparison of our data to the traditional gaze tracking methods. Lastly we introduce a video data set together with our gaze data.

\subsection{Gaze Tracking Data Collection}
\label{collection}

As our gaze data collection method is very flexible we deploy it on the Internet. We have built a Flash-based application that instructs the user, shows the required number of tutorials and then displays a batch of videos followed by a triplets chart. To ensure the proper size of the video we disallow participation of users with screen resolution less than $1024\times 768$. We additionally request the browser window to be maximized on the display. We use batches of 6 videos in every session. To reach as many participants as possible we employ Amazon Mechanical Turk. Since we have a variety of different videos and different frames of interest we allow each user to perform the entire session several times. The participants are paid US \$0.15 for each completed session. Thus collecting 100 gaze locations for a single video frame would cost US \$2.50.

To make the self-reported gaze location more accurate we employ several different techniques: First, the triplets of letters and numbers are used to explicitly identify gaze location and reject erroneous data entry.
Second, we show the chart for a short period of time to minimize eye movement after the video ends. Third, we use tutorials with approval radius for participant screening.

\begin{figure}[tbh]
    \centering
    \includegraphics[width=0.8\linewidth]{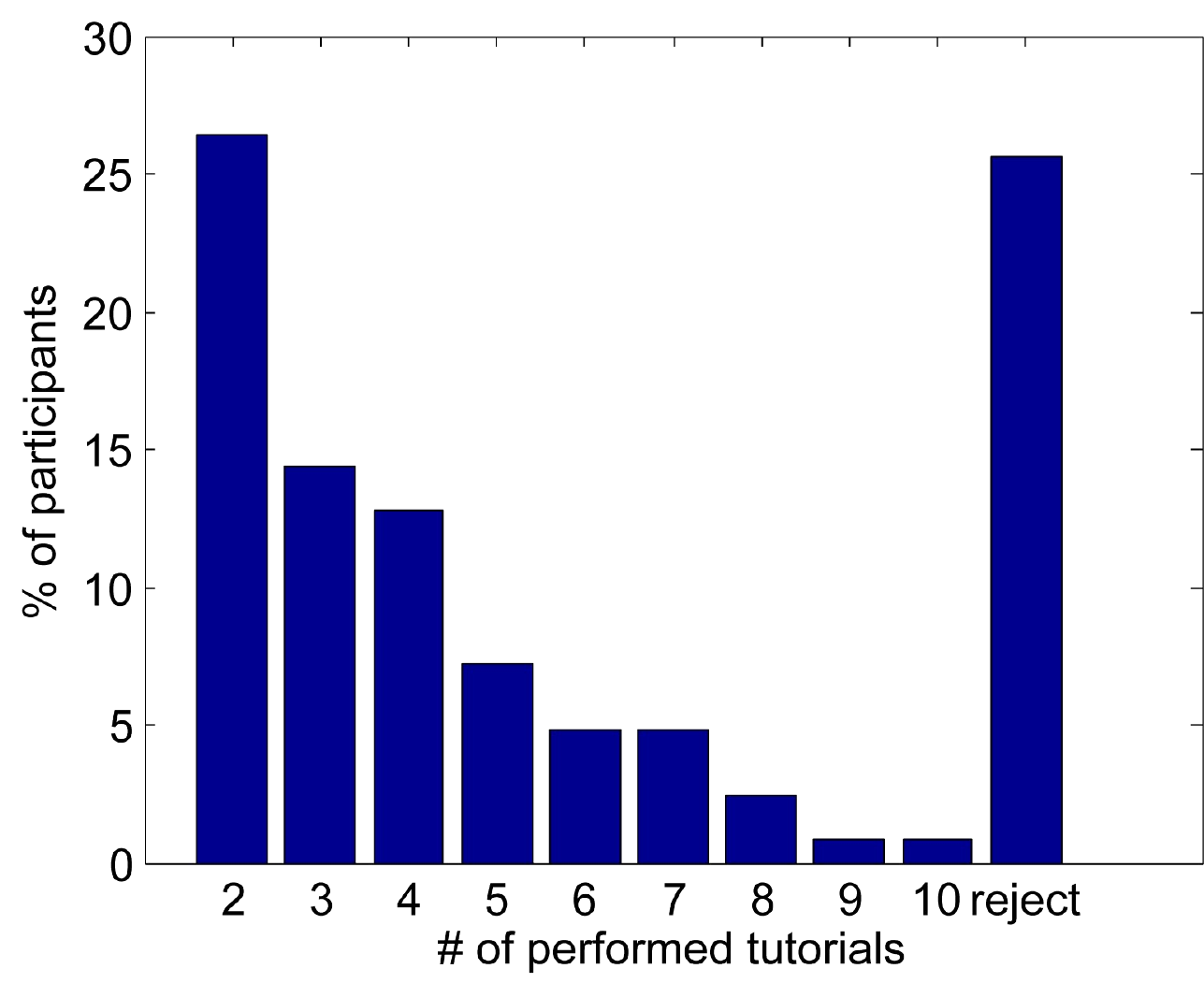}
  \caption{\label{fig:tut_stat}
Statistics of the tutorials. One quarter of the participants were rejected for performing too many tutorials. The others have managed to complete the required two tutorials. Only about one quarter of users managed not to fail any tutorial.
}
\end{figure}

The importance of the tutorial is emphasized by the success statistics. Only about quarter of the participants succeeded in passing the first two tutorials. The other half needed more trials in order to pass two. Additionally, about one quarter of the participants were rejected for exceeding the maximum number of allowed tutorials (10). The tutorial statistics are illustrated in Figure~\ref{fig:tut_stat}.

All the participants who succeeded in two tutorials were allowed to continue to the videos. After each video the user is requested to report the triplet he noted on the screen. If the reported triplet exists, the current video experiment is defined as successful. We use 6 random videos and frames of interest in every batch.
Those participants who did complete the tutorials had a very high success rate -- over 95\% on average.

\subsection{Comparison to the Traditional Gaze Tracking}
\label{compare}

The first experiment we performed is validation of the gaze location data collected using our method versus traditional gaze tracking. We used a publicly available gaze-tracked video dataset called DIEM (Dynamic Images and Eye Movements)~\cite{mital2010clustering}. This dataset includes 84 high-definition videos from different styles, such as movie trailers, ads, sport events etc. However, most of the videos are professionally produced.

For validation we choose three random frames from four videos. The frames are approximately evenly spread across each video, and the videos represent several different genres from the original corpus. Using our gaze data collection method we gather about 50 gaze locations per frame. The cost of this experiment is about US \$20. We compare this data to the hardware gaze tracking data for the same frames, provided by the DIEM data set.

\begin{figure}[htb]
    \centering
    \includegraphics[width=0.8\linewidth]{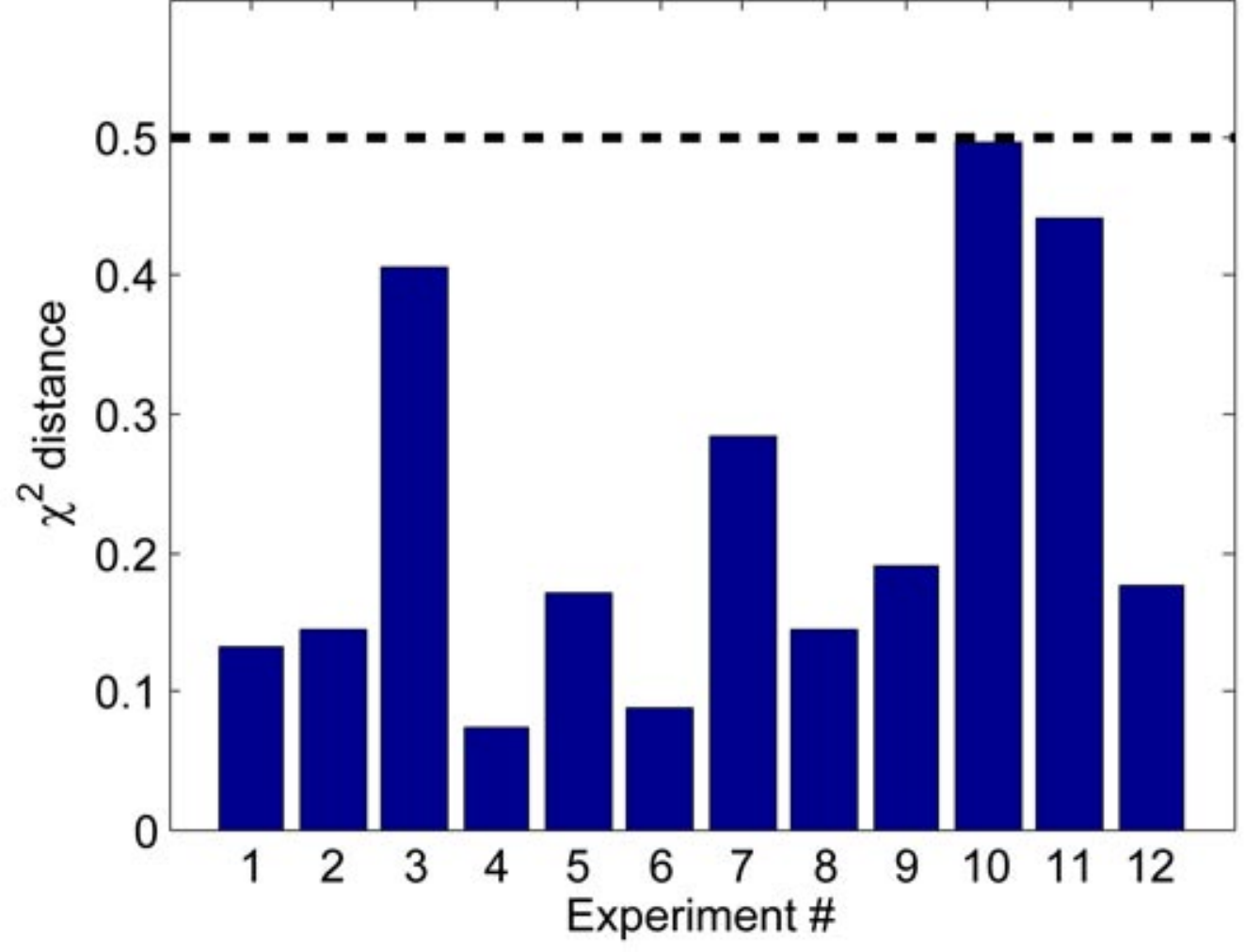}
  \caption{\label{fig:diem_chi-sq}
Comparison of our data versus DIEM's gaze tracking. $\chi^2$ distance measures the distance between the distribution estimated form our gaze location data and from gaze tracking. The dashed line indicates the distance to a uniform distribution of points over the frame. As one can see, our data is close to the traditional gaze tracking.
}
\end{figure}

\begin{figure*}[tb]
    \centering
    \begin{tabular}{cc}
		Our data & Traditional gaze tracking \\
        \includegraphics[height=3.9cm]{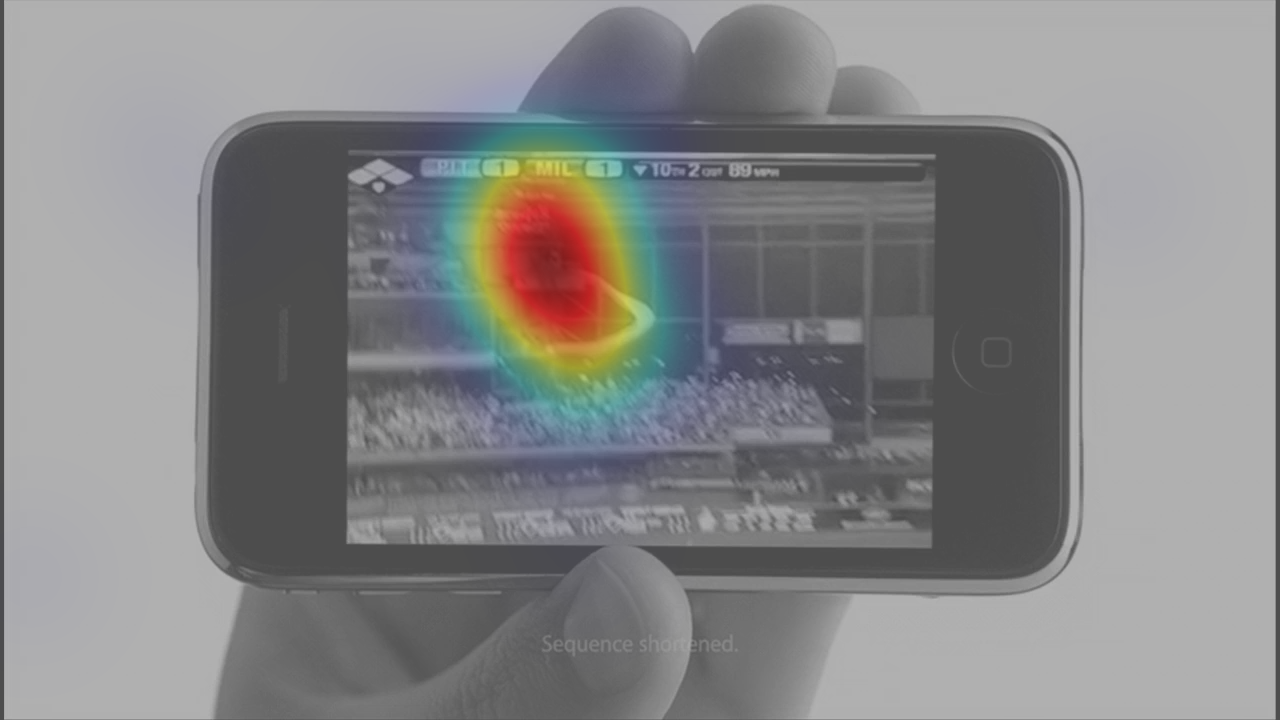} &
		\includegraphics[height=3.9cm]{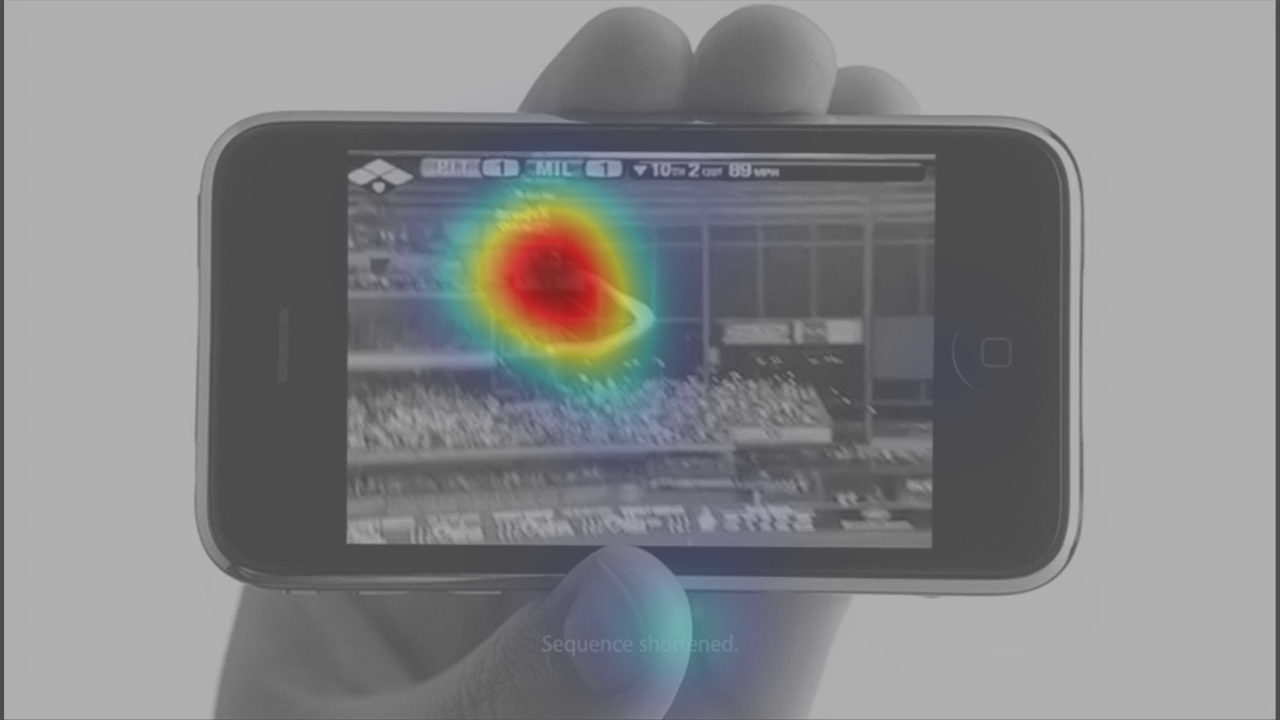} \\
        	 \includegraphics[height=3.9cm]{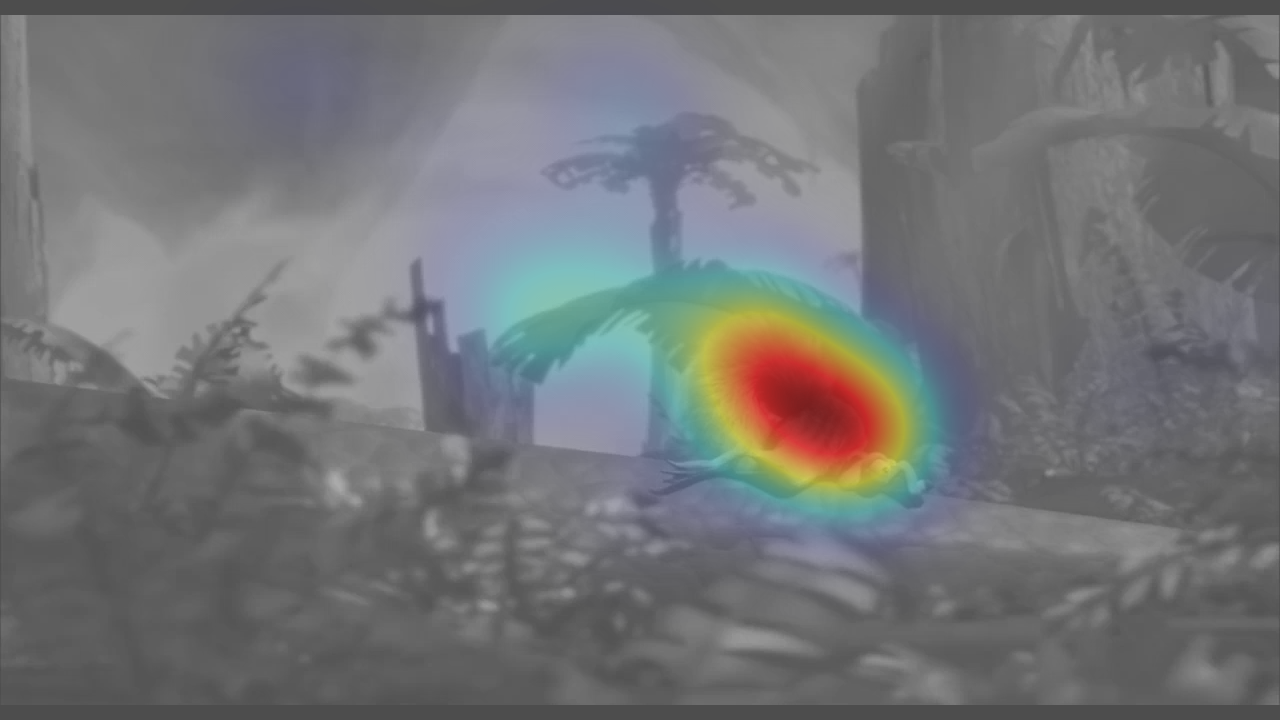} &
		 \includegraphics[height=3.9cm]{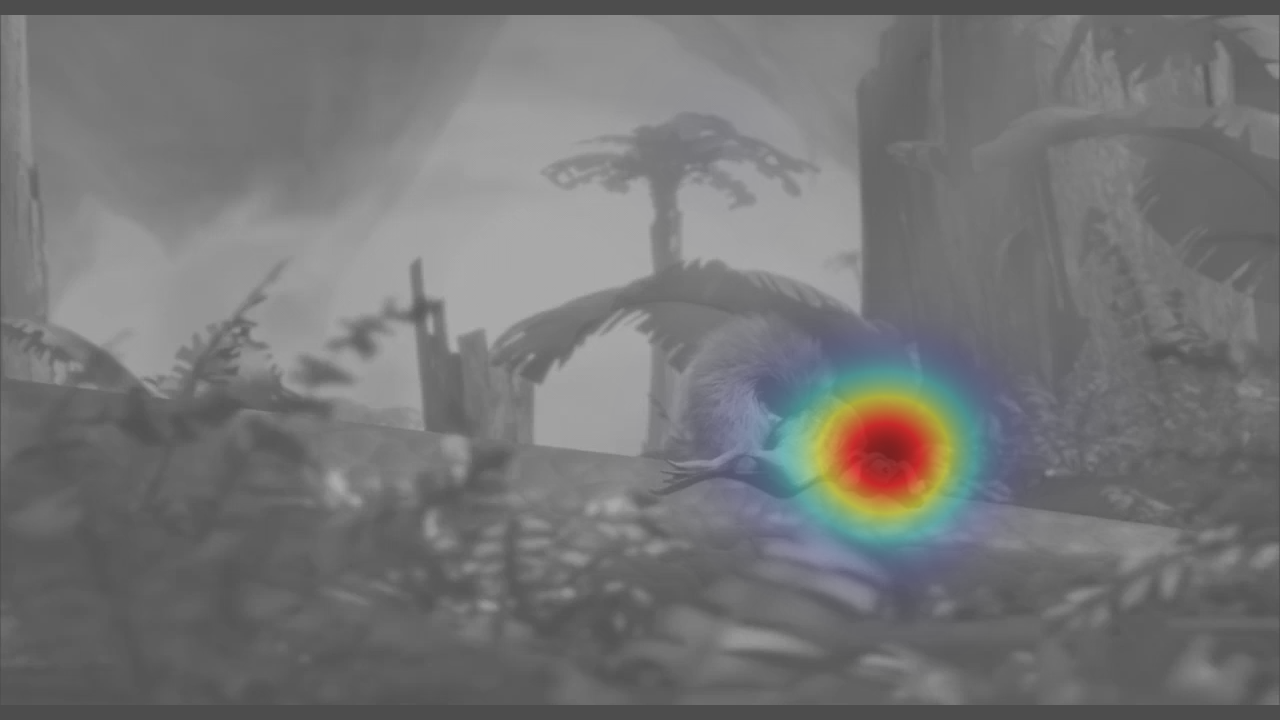} \\
		(a) & (b)
    \end{tabular}
  \caption{\label{fig:diem_vis}
Gaze probability estimated from our data (a) and from gaze tracking (b). Top frame corresponds to the column 5 in Figure~\ref{fig:diem_chi-sq}, and bottom to the column 3. It can be seen that when the distance is high (bottom) the distribution estimated from our data still indicates the correct place but less accurately.
}
\end{figure*}

\begin{figure*}[tbh]
\centering
\begin{tabular}{cccc}
\hline
 & Frame of interest & Our gaze location & DIEM gaze tracking \\
\hline
\\
\raisebox{1.4cm}{(1)} &
\includegraphics[width=5.0cm]{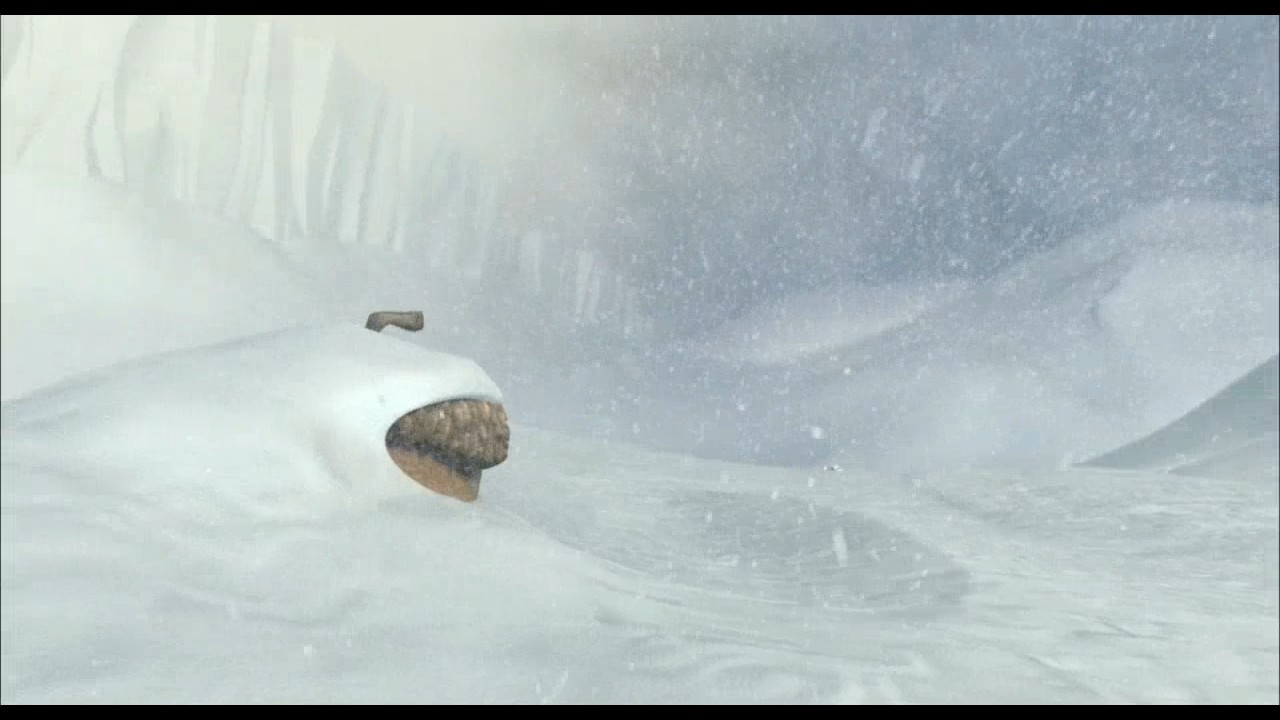} &
\includegraphics[width=5.0cm]{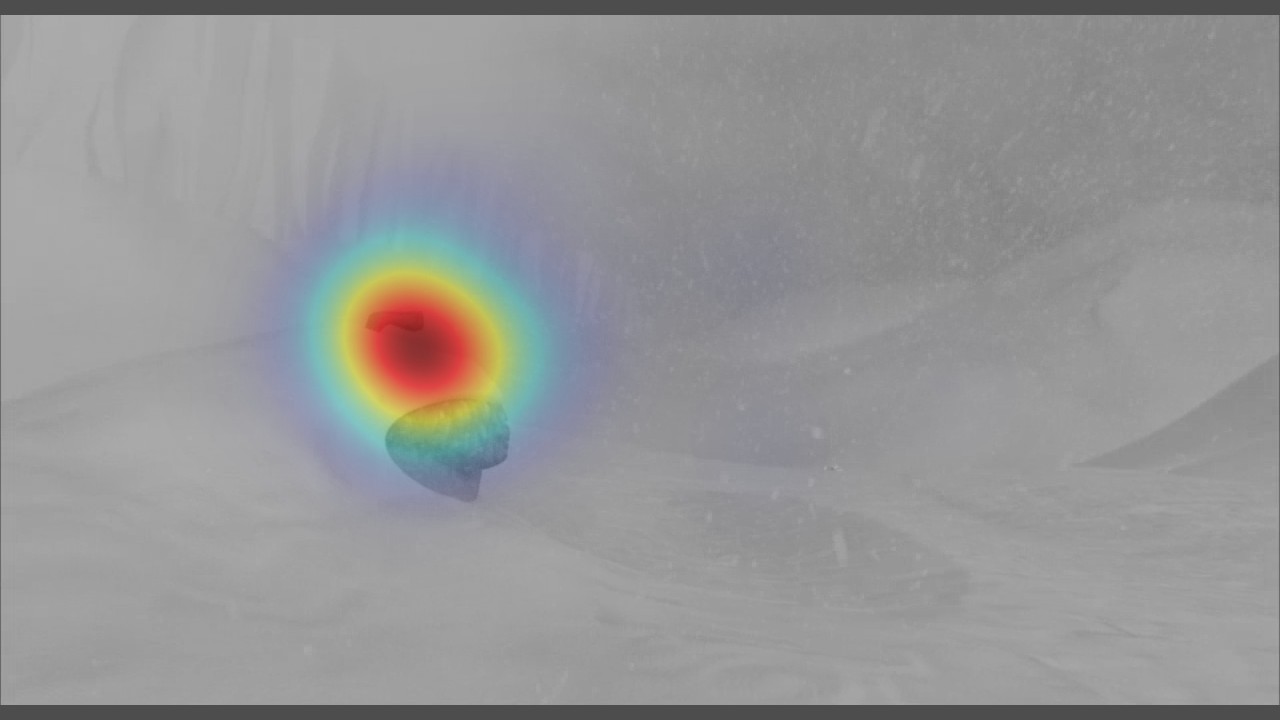} &
\includegraphics[width=5.0cm]{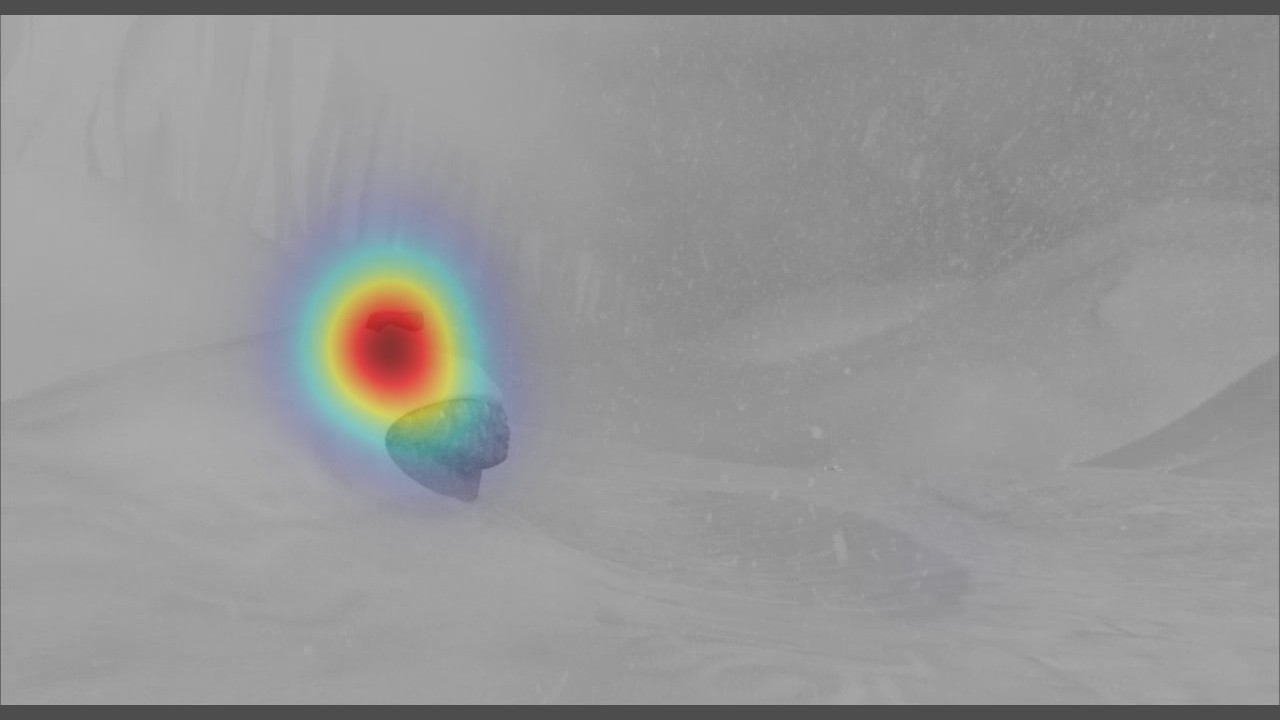} \\
\raisebox{1.4cm}{(2)} &
\includegraphics[width=5.0cm]{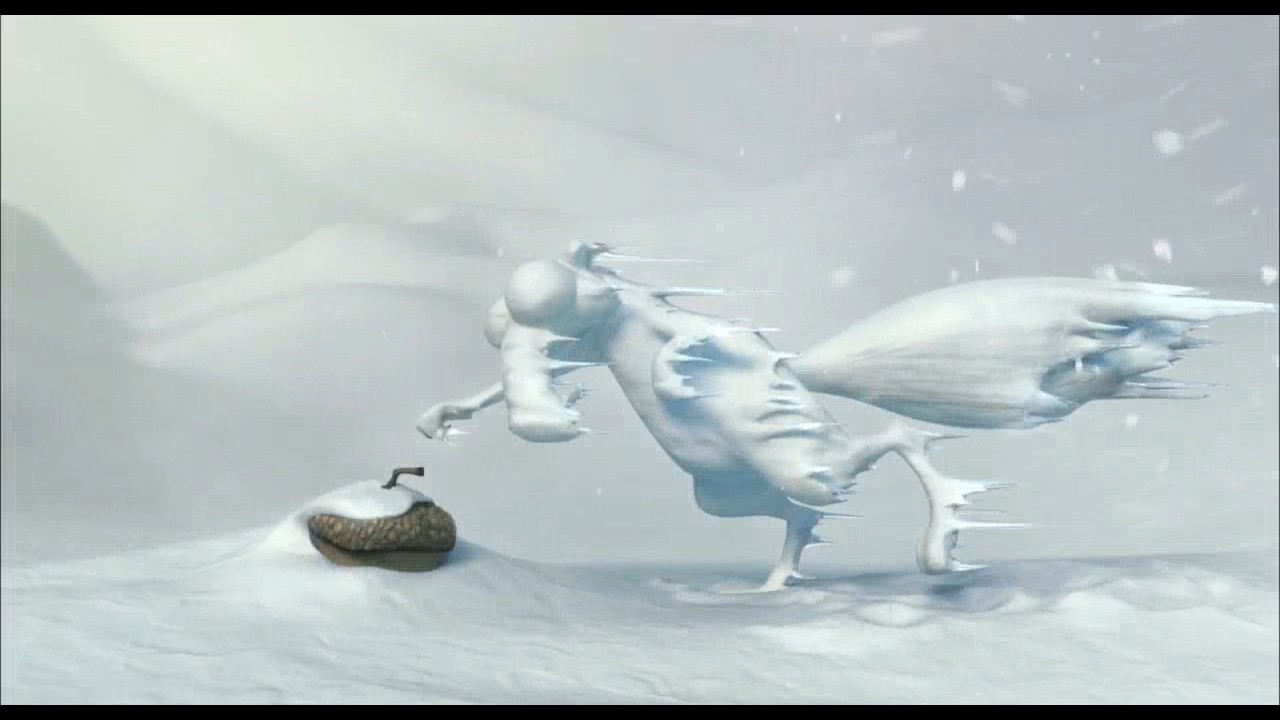} &
\includegraphics[width=5.0cm]{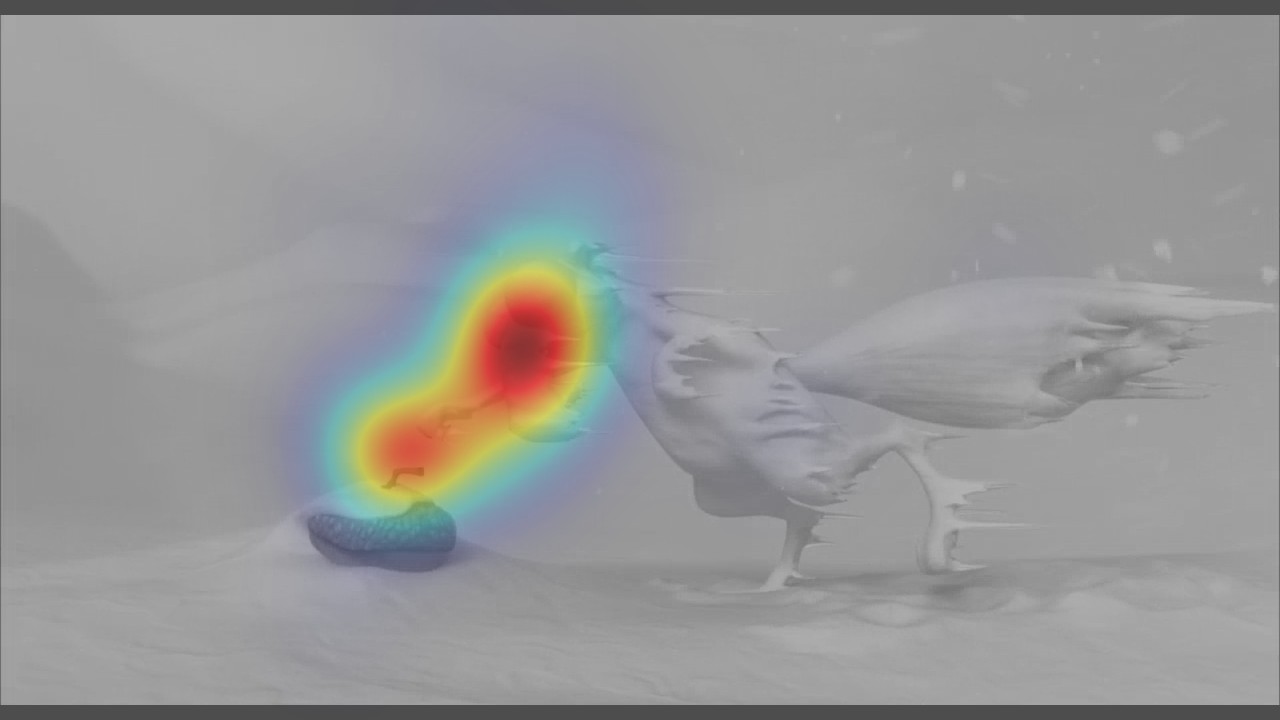} &
\includegraphics[width=5.0cm]{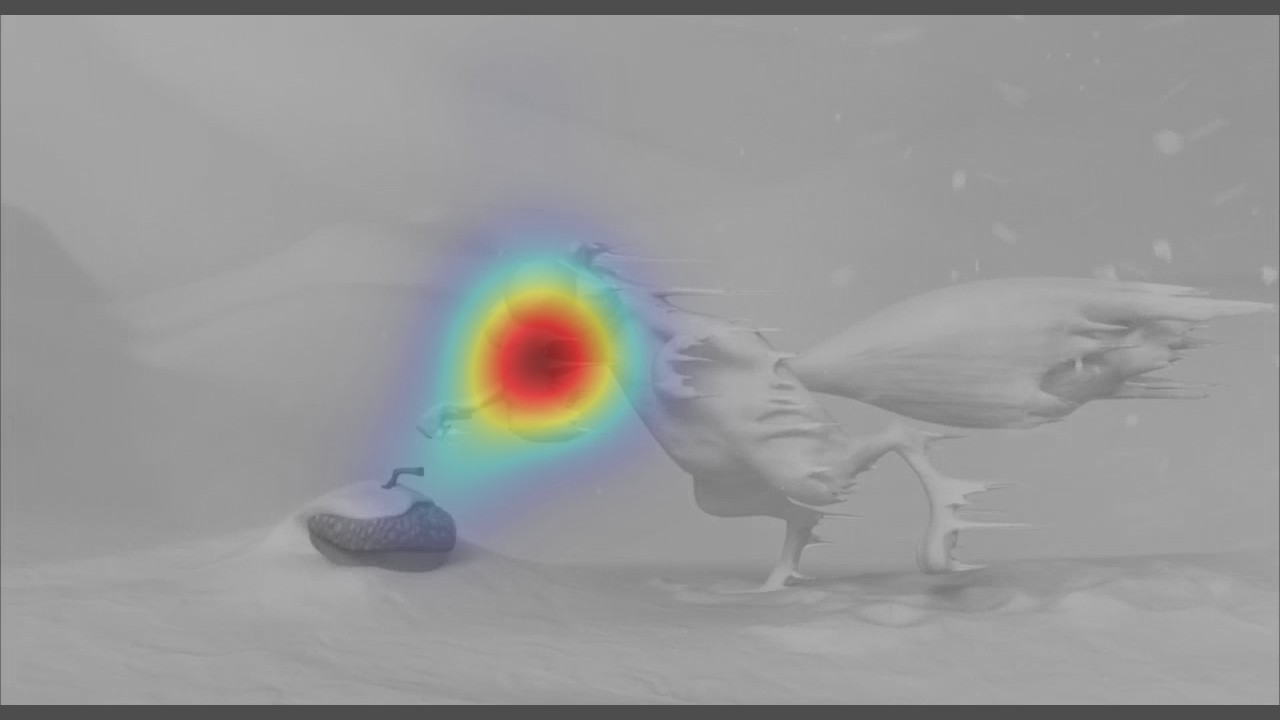} \\
\raisebox{1.4cm}{(3)} &
\includegraphics[width=5.0cm]{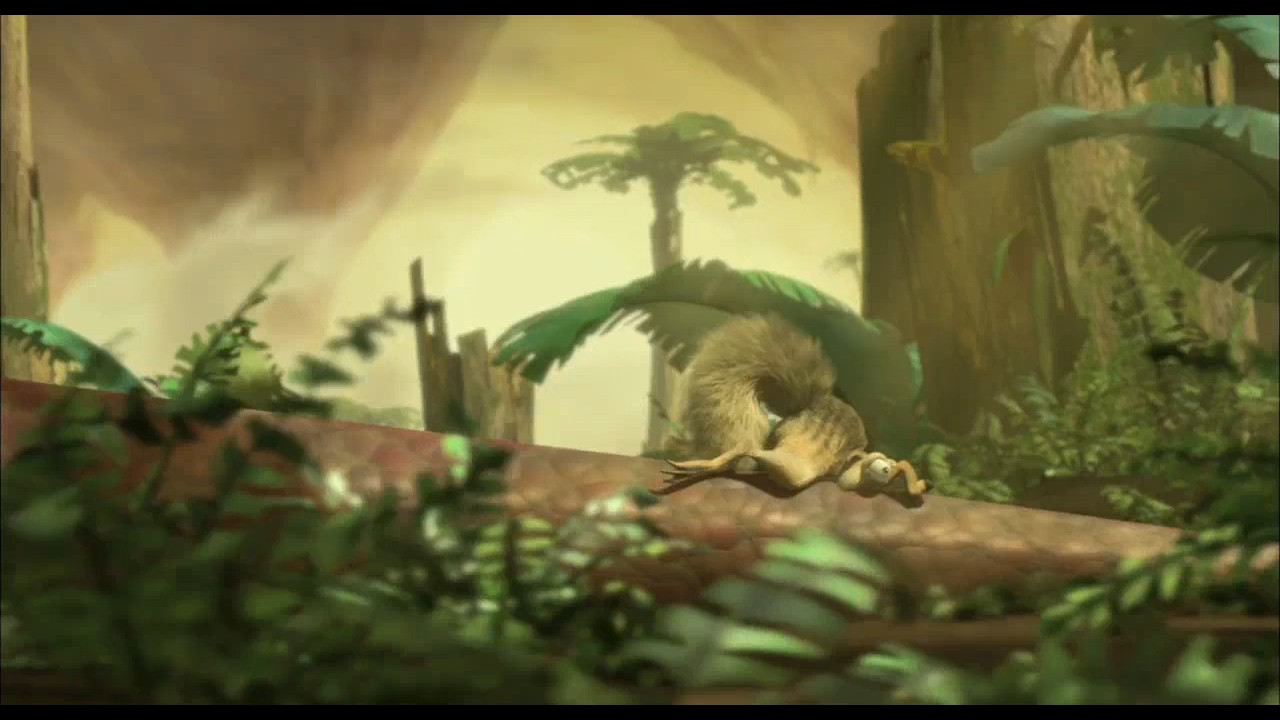} &
\includegraphics[width=5.0cm]{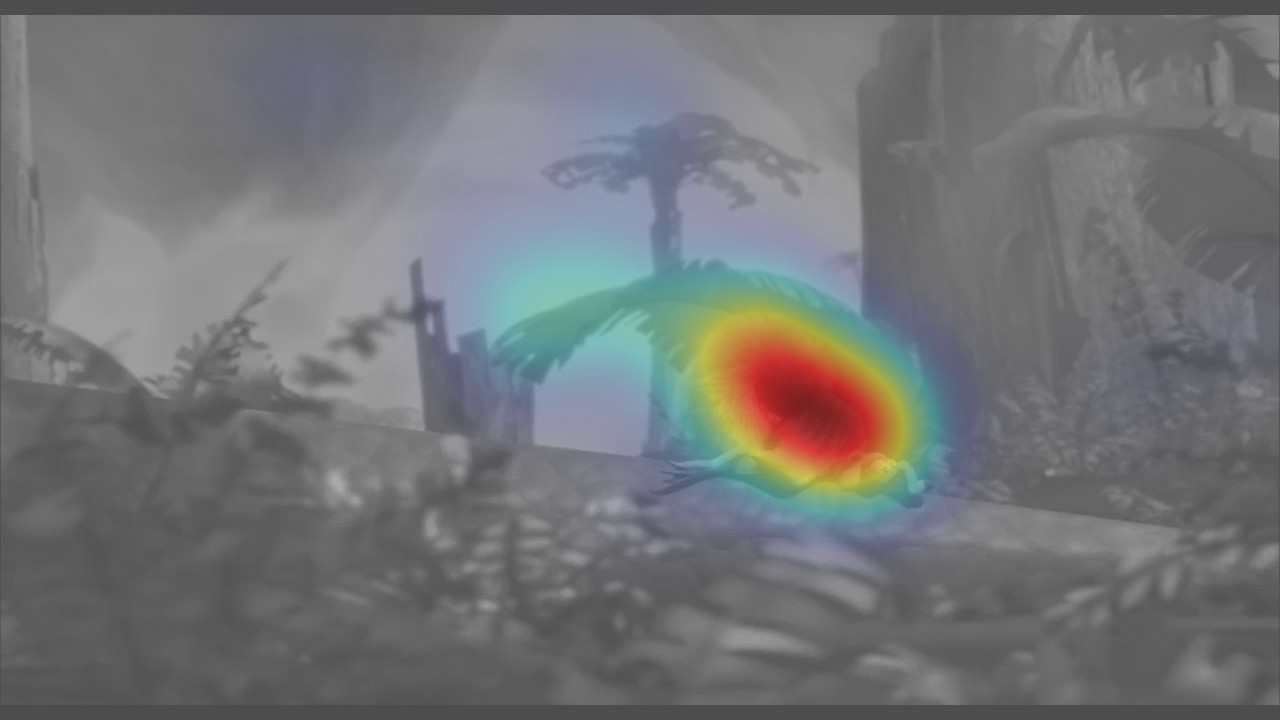} &
\includegraphics[width=5.0cm]{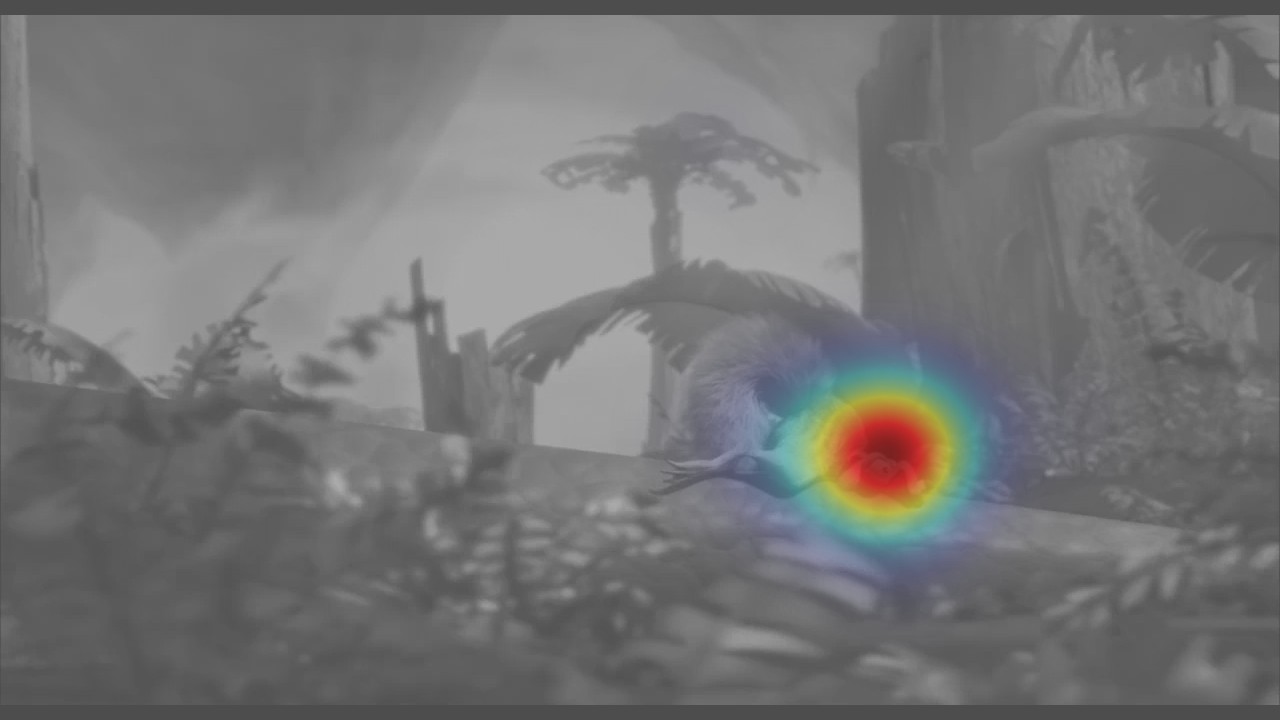} \\
\raisebox{1.4cm}{(4)} &
\includegraphics[width=5.0cm]{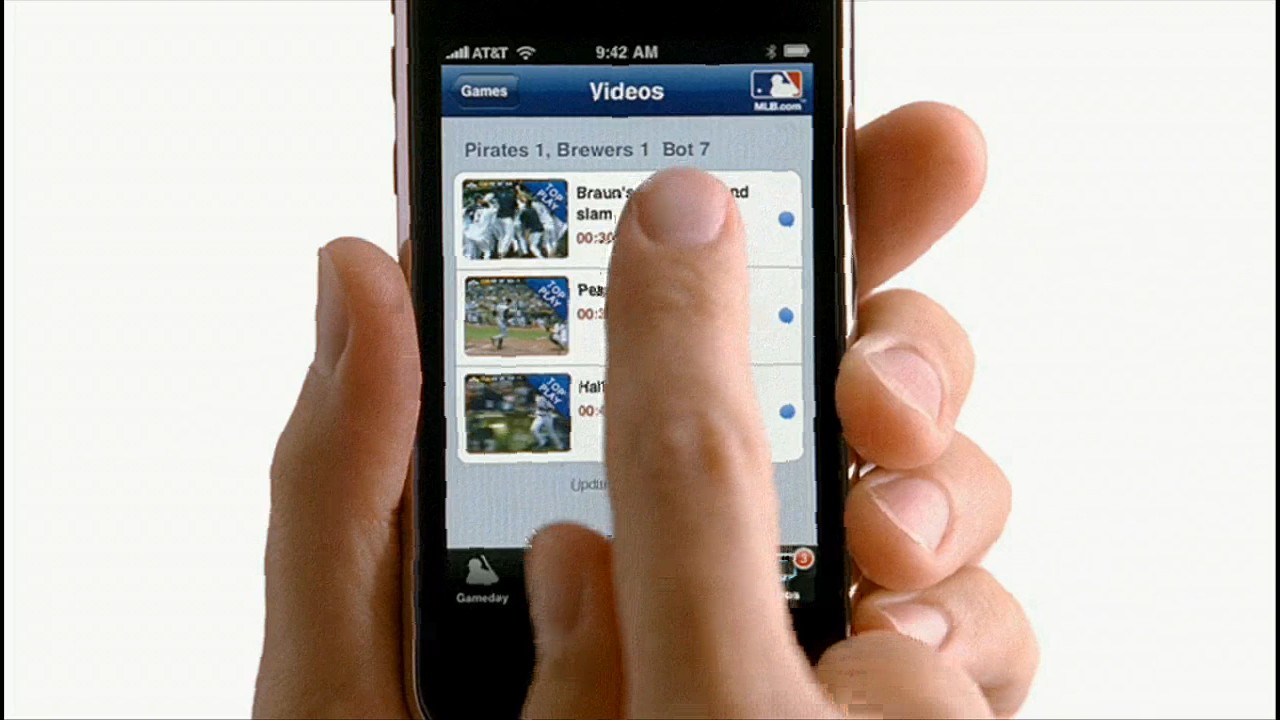} &
\includegraphics[width=5.0cm]{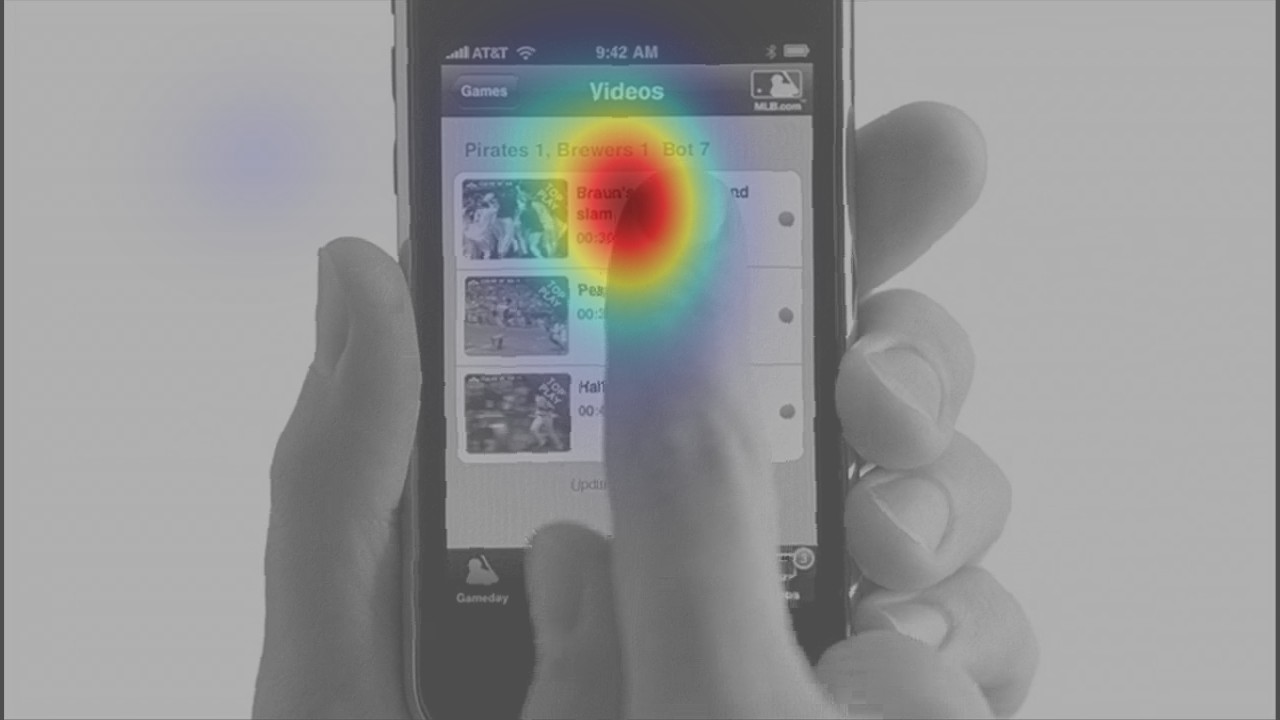} &
\includegraphics[width=5.0cm]{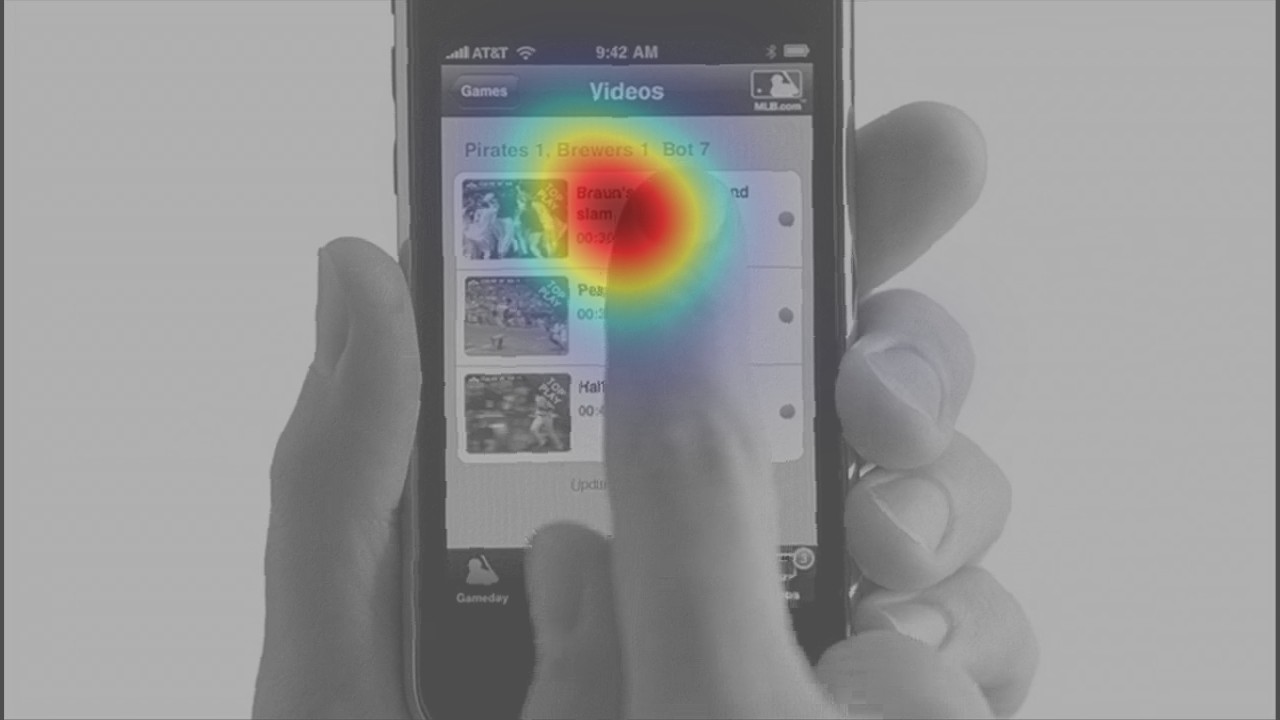} \\
\raisebox{1.4cm}{(5)} &
\includegraphics[width=5.0cm]{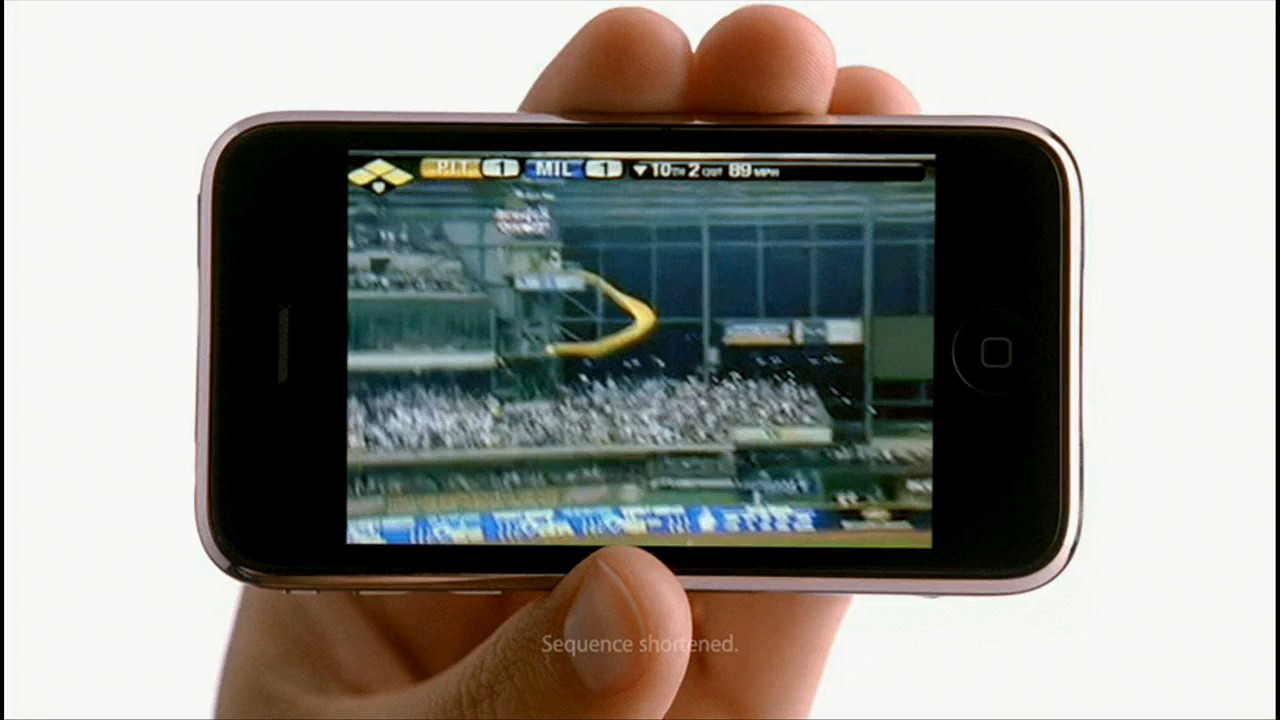} &
\includegraphics[width=5.0cm]{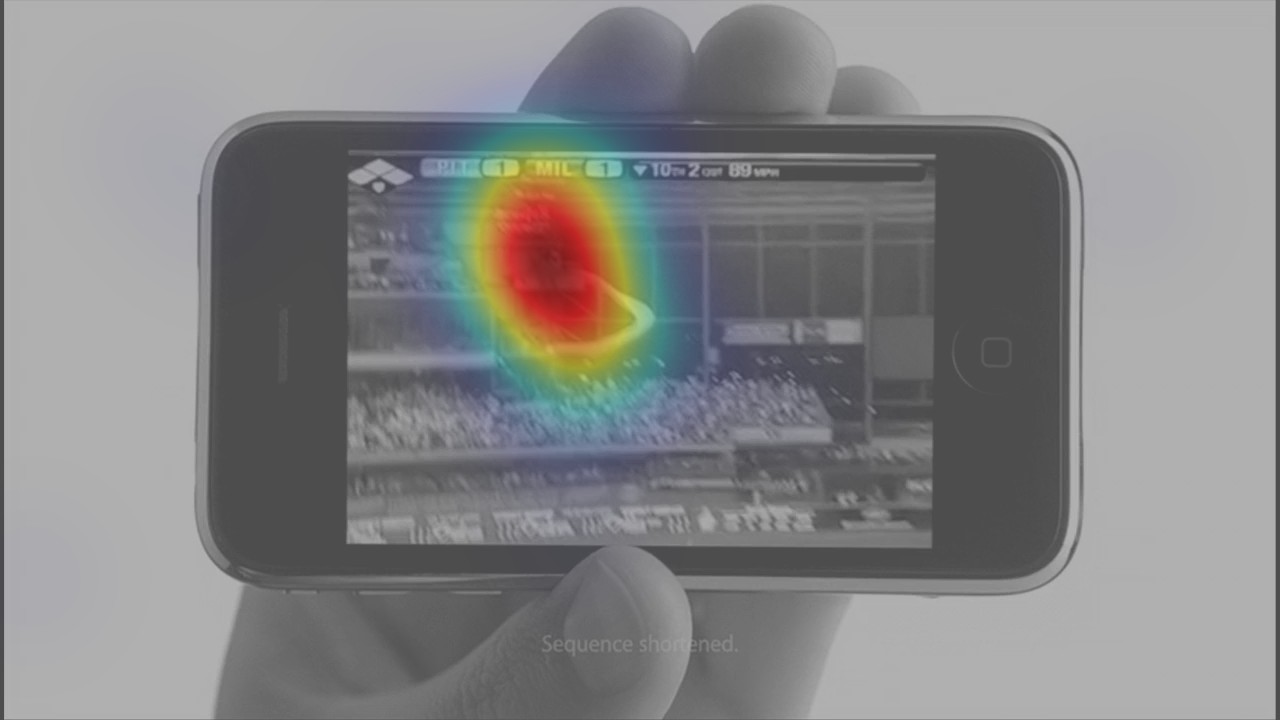} &
\includegraphics[width=5.0cm]{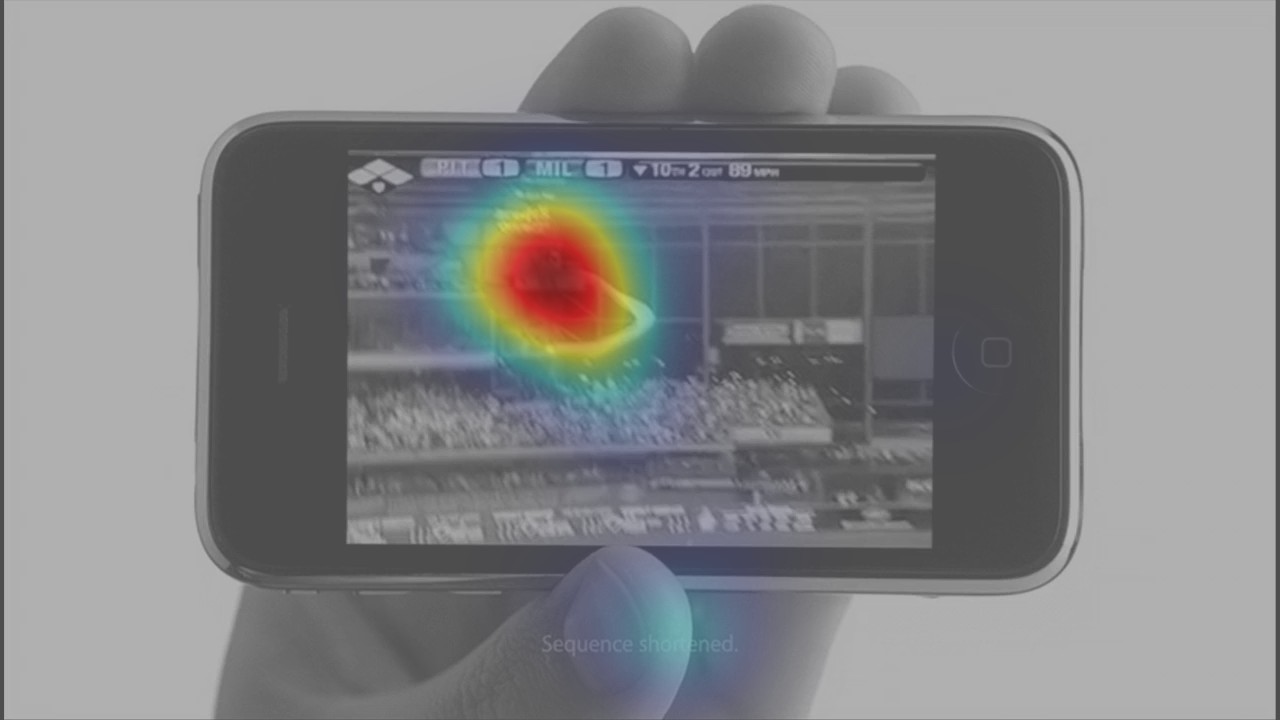} \\
\raisebox{1.4cm}{(6)} &
\includegraphics[width=5.0cm]{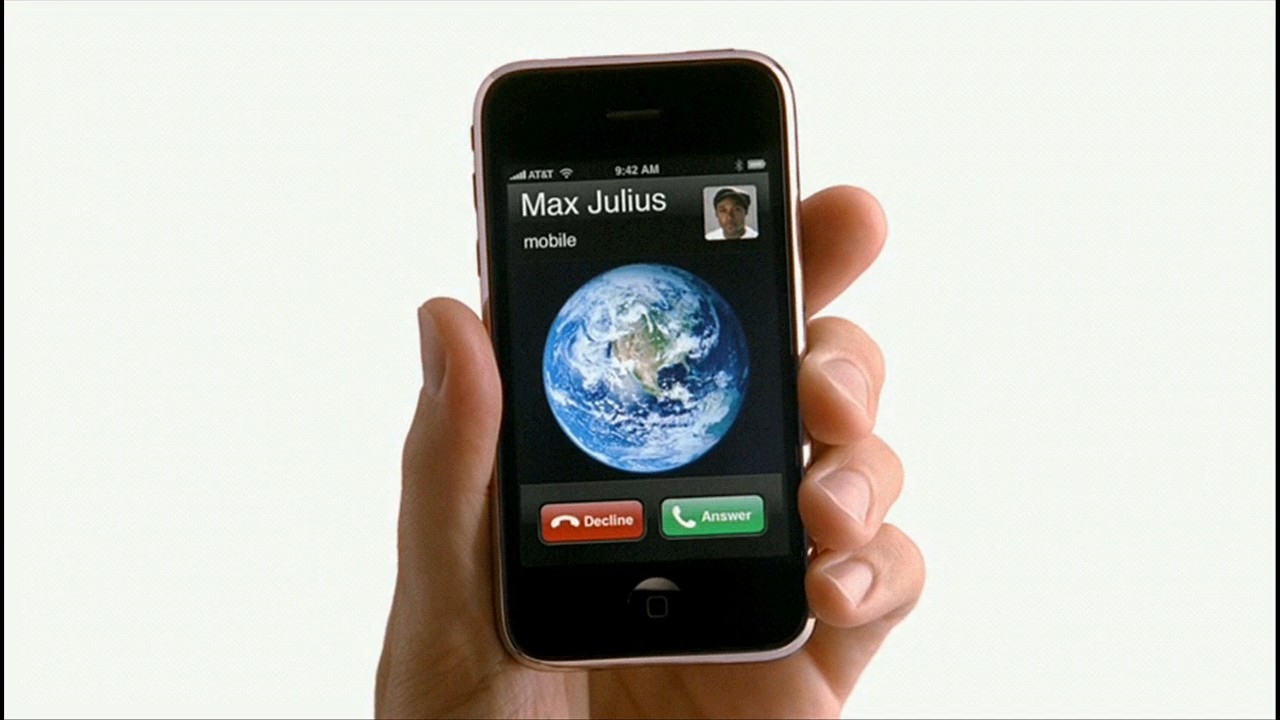} &
\includegraphics[width=5.0cm]{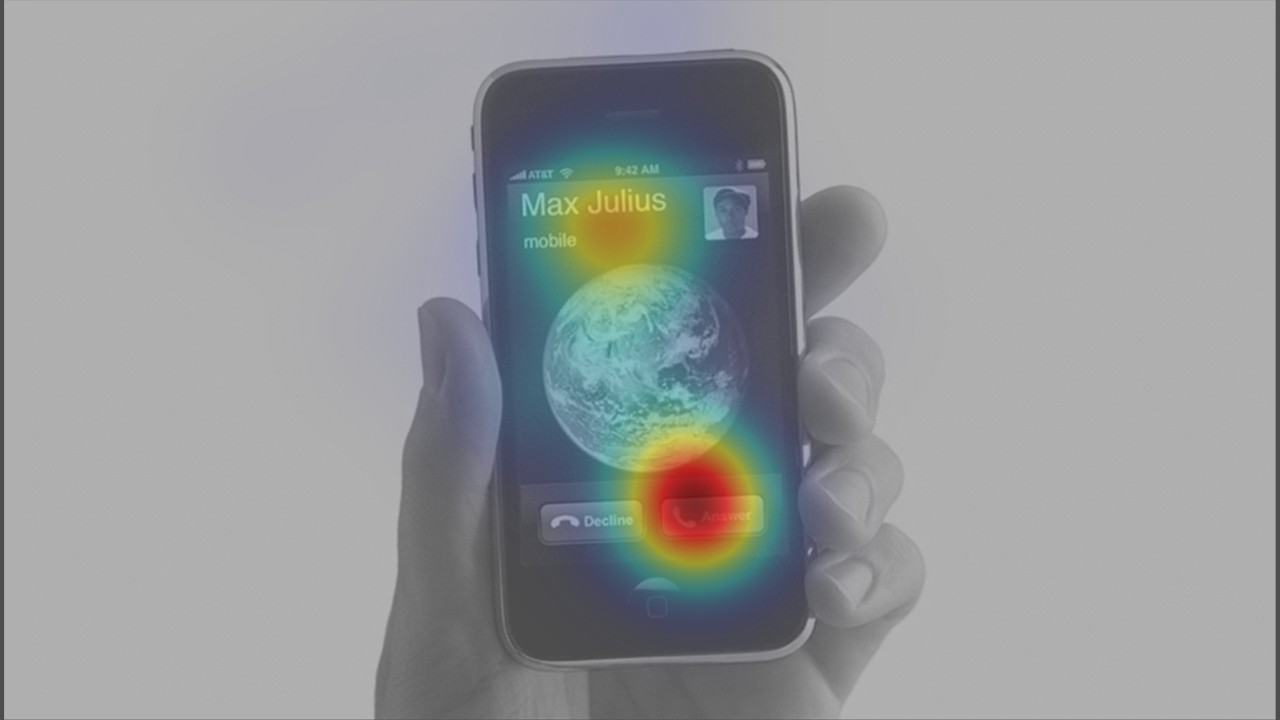} &
\includegraphics[width=5.0cm]{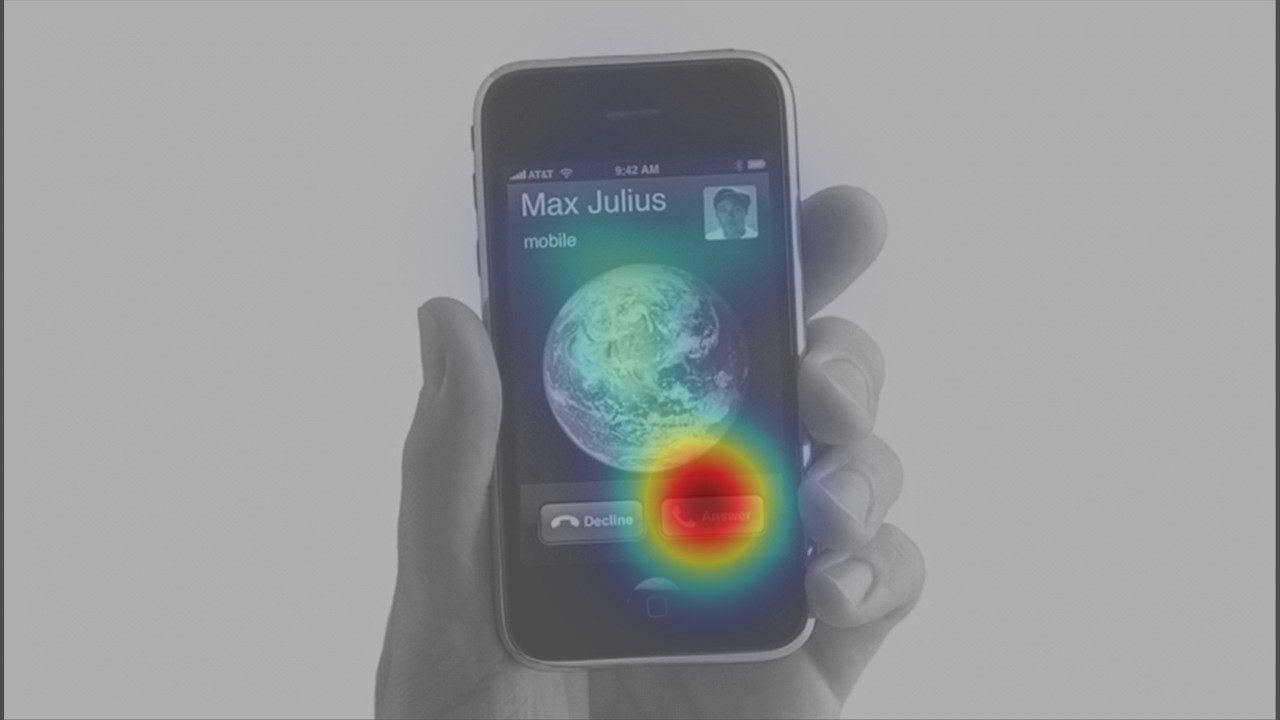} \\
\raisebox{1.4cm}{(7)} &
\includegraphics[width=5.0cm]{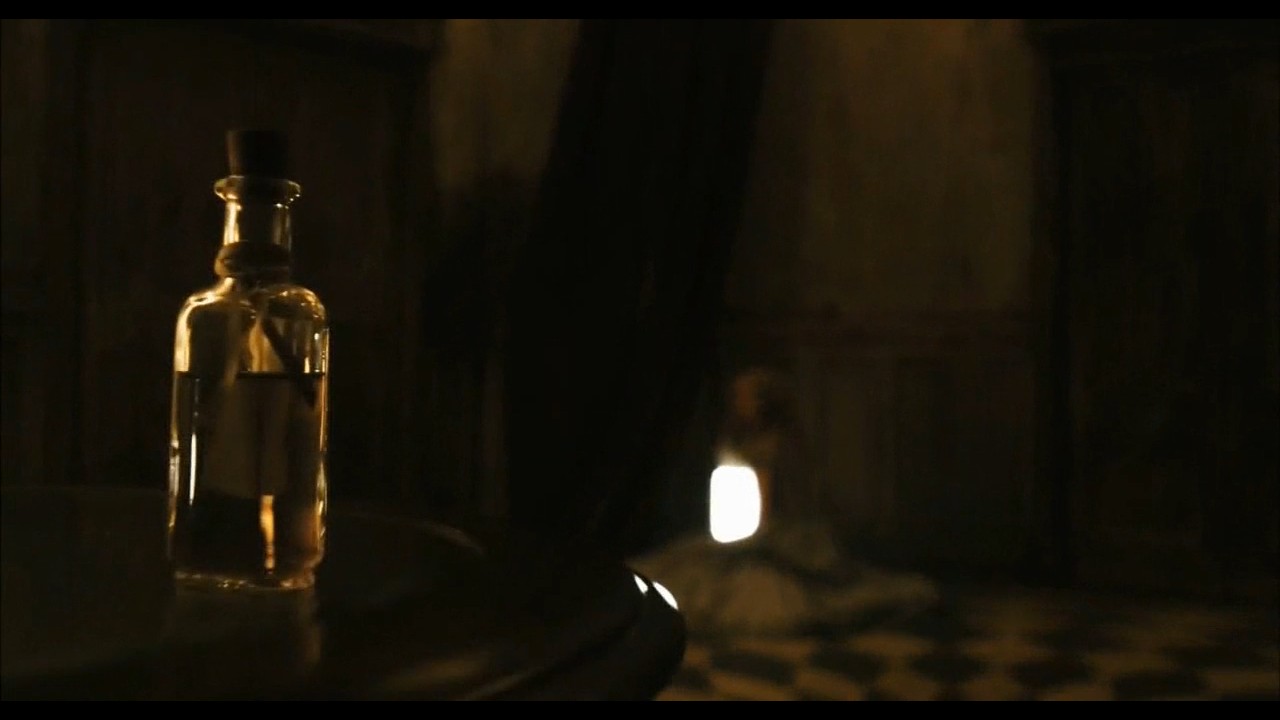} &
\includegraphics[width=5.0cm]{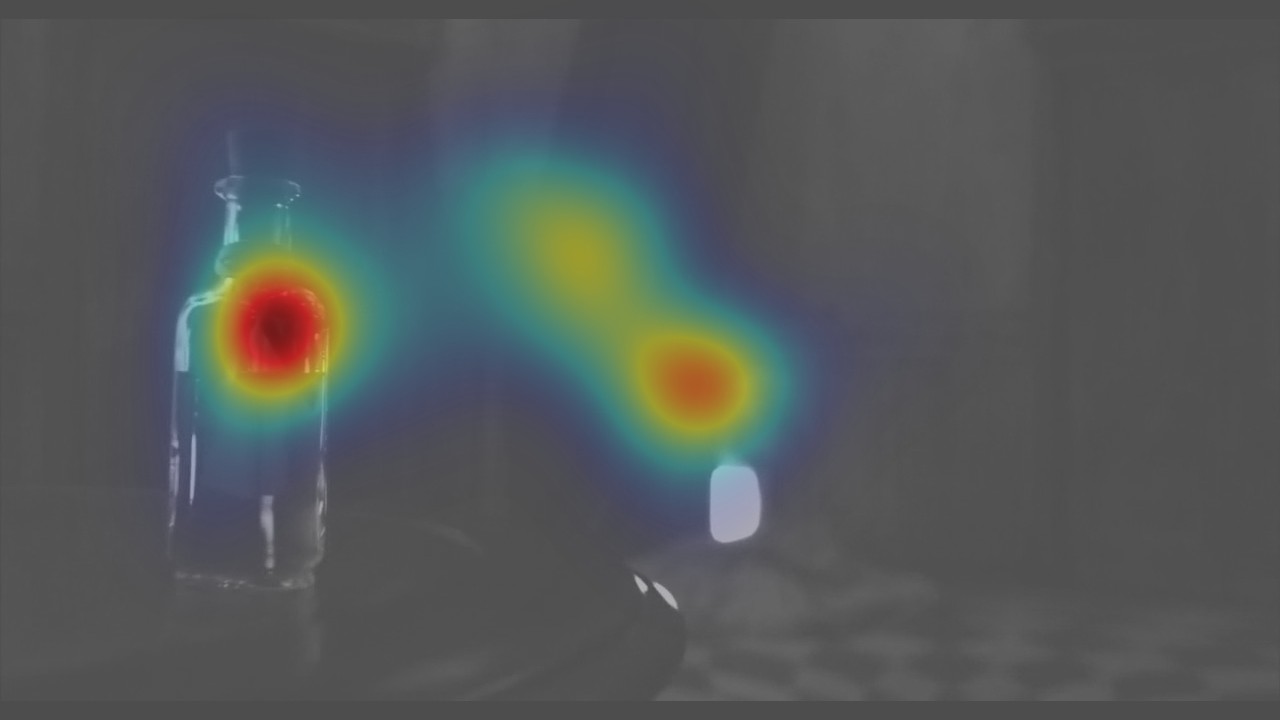} &
\includegraphics[width=5.0cm]{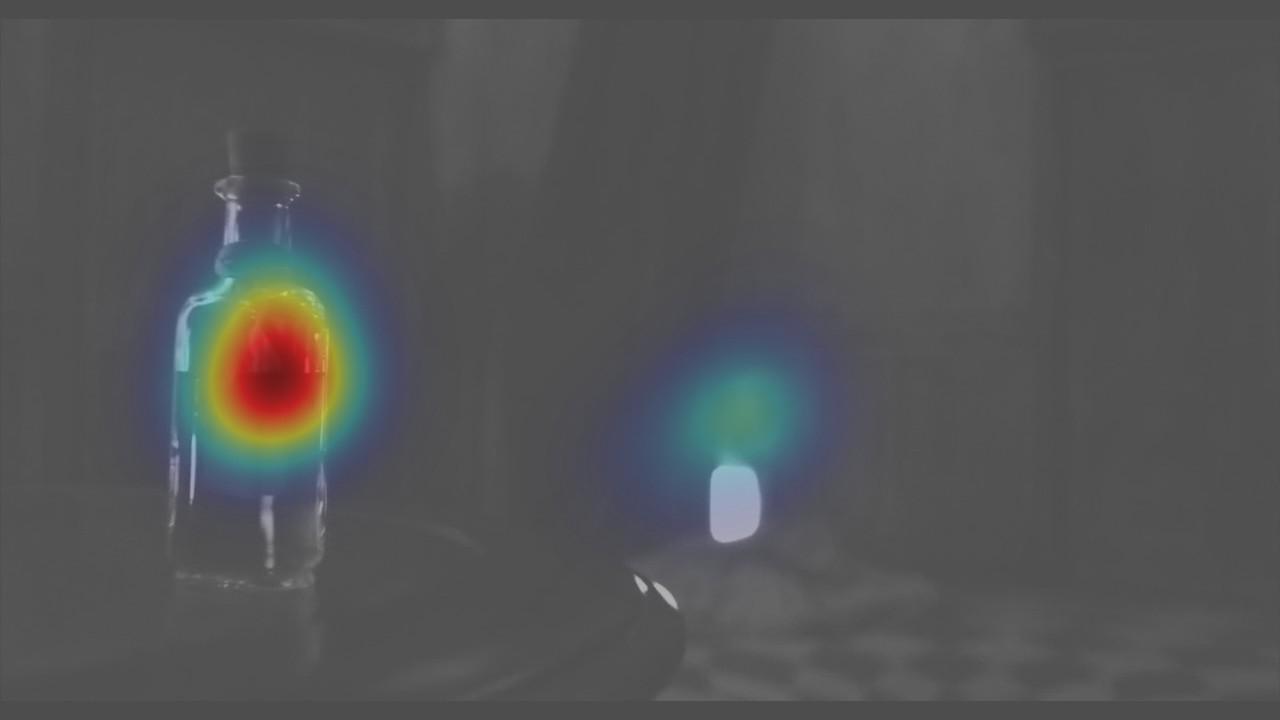} \\
\end{tabular}
  \caption{\label{fig:diem_vis_all}
  Additional results for gaze probability estimation. Left to right: original frame, our gaze location data and DIEM's gaze tracking results. The numbers correspond to Figure~\ref{fig:diem_chi-sq}.
}
\end{figure*}

\begin{figure*}[htb]
\centering
\begin{tabular}{cccc}
\hline
 & Frame of interest & Our gaze location & DIEM gaze tracking \\
\hline
\\
\raisebox{1.4cm}{(8)} &
\includegraphics[width=5.0cm]{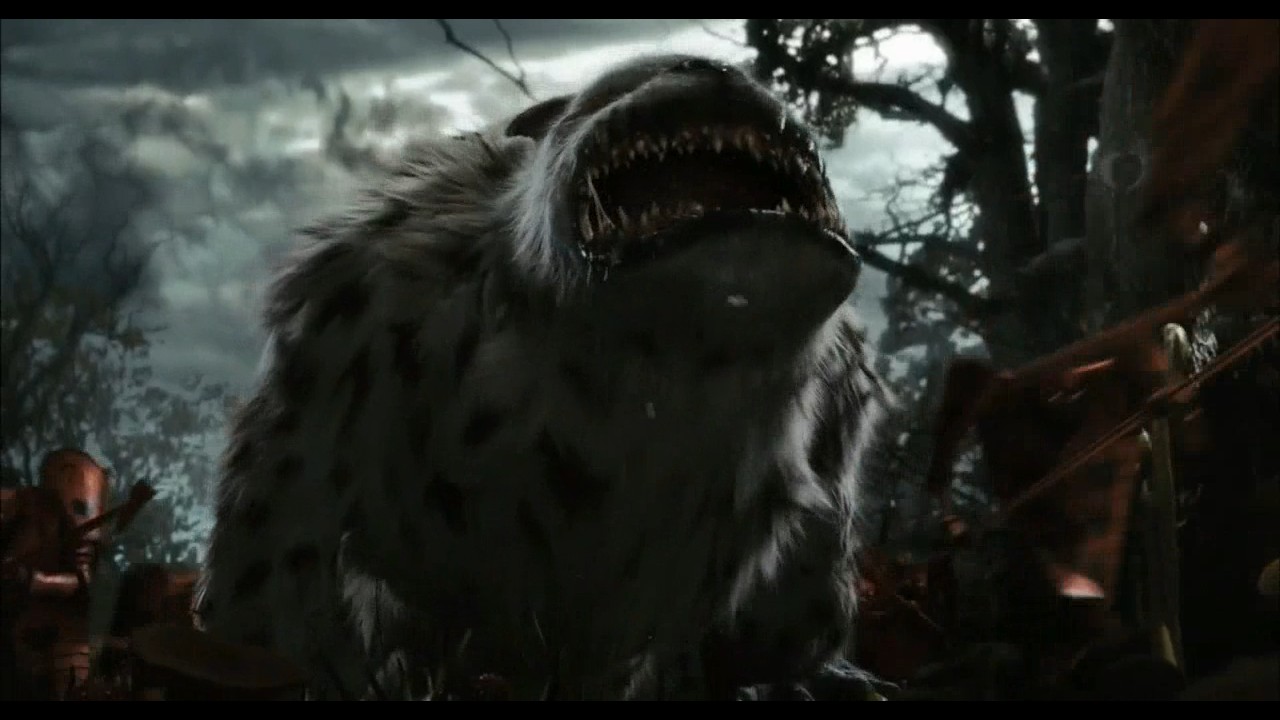} &
\includegraphics[width=5.0cm]{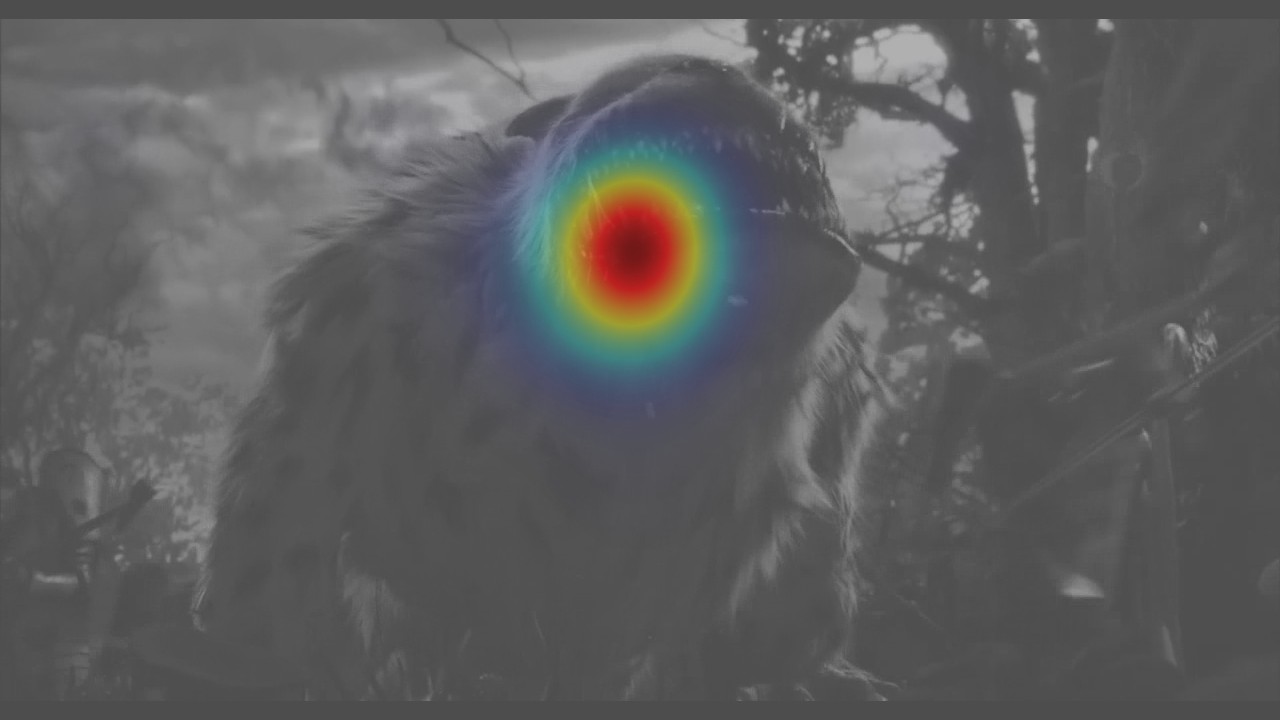} &
\includegraphics[width=5.0cm]{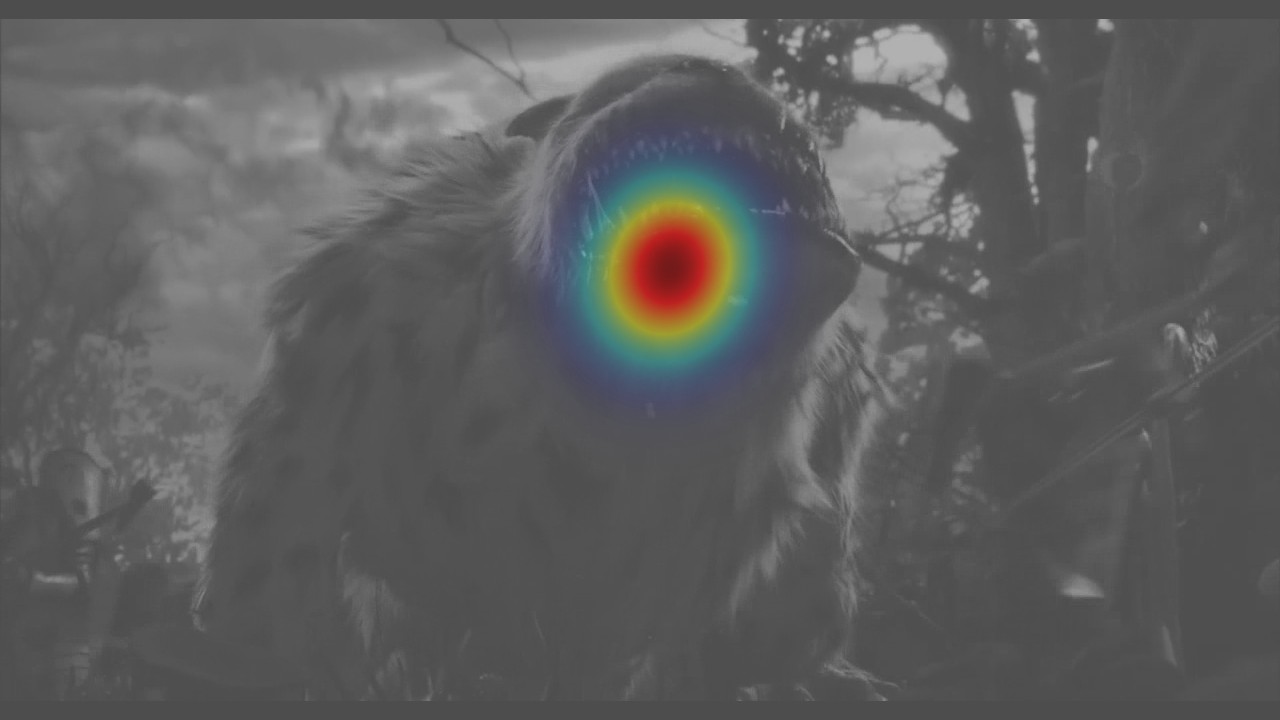} \\
\raisebox{1.4cm}{(9)} &
\includegraphics[width=5.0cm]{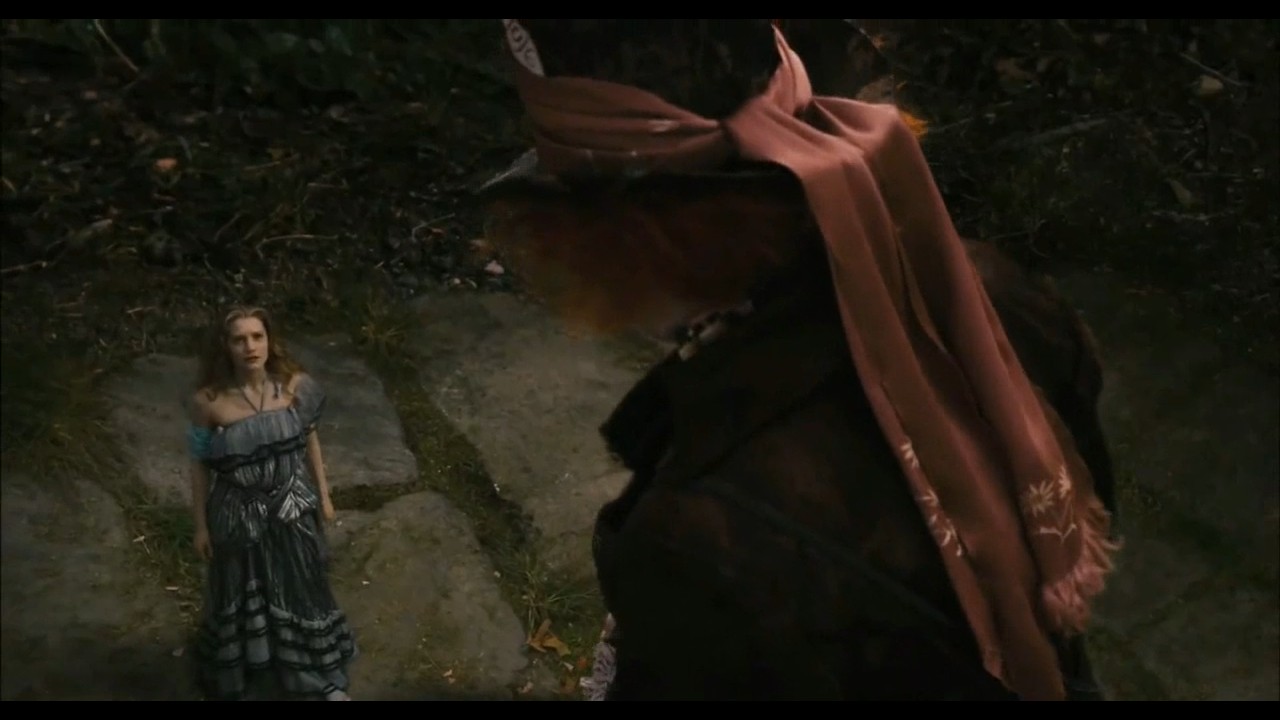} &
\includegraphics[width=5.0cm]{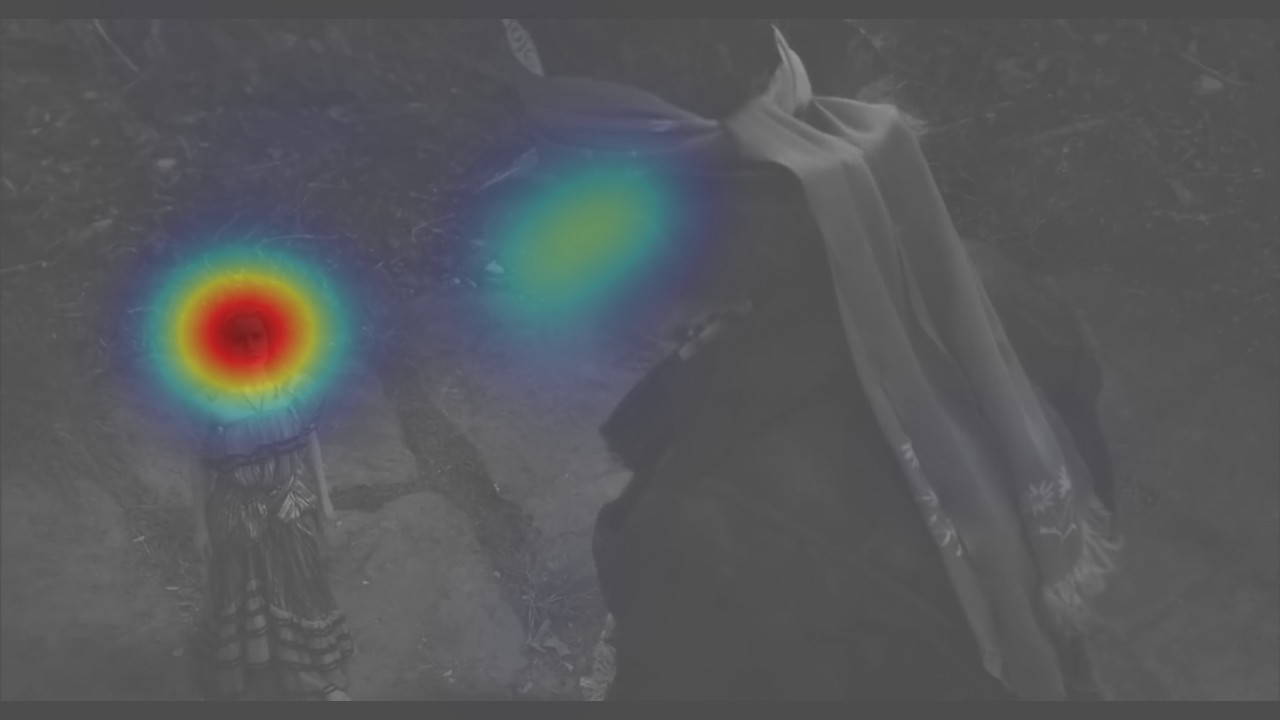} &
\includegraphics[width=5.0cm]{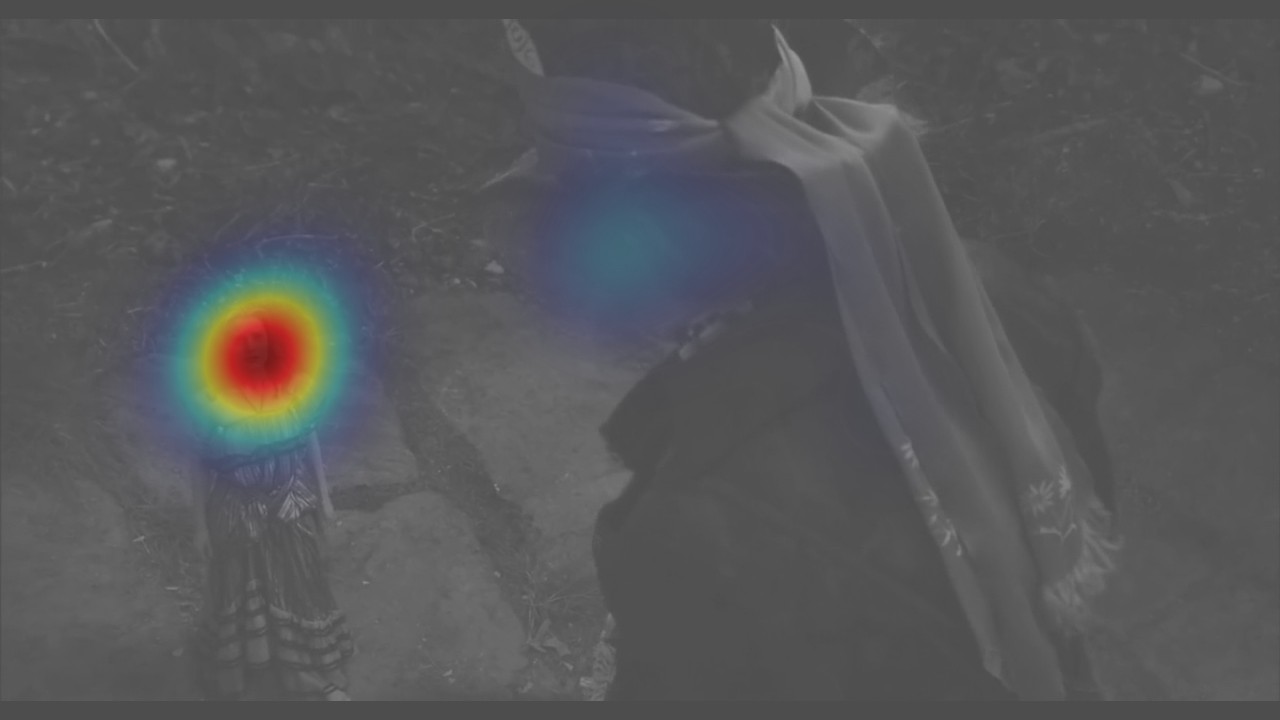} \\
\raisebox{1.4cm}{(10)} &
\includegraphics[width=5.0cm]{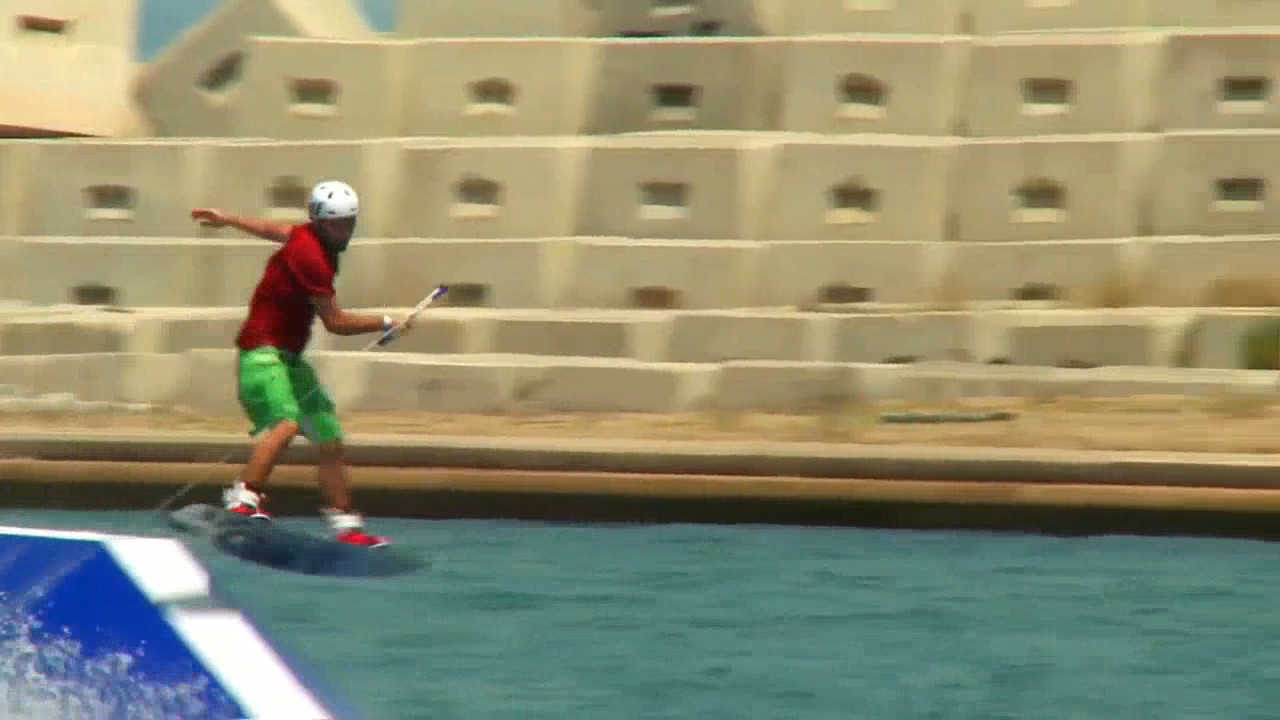} &
\includegraphics[width=5.0cm]{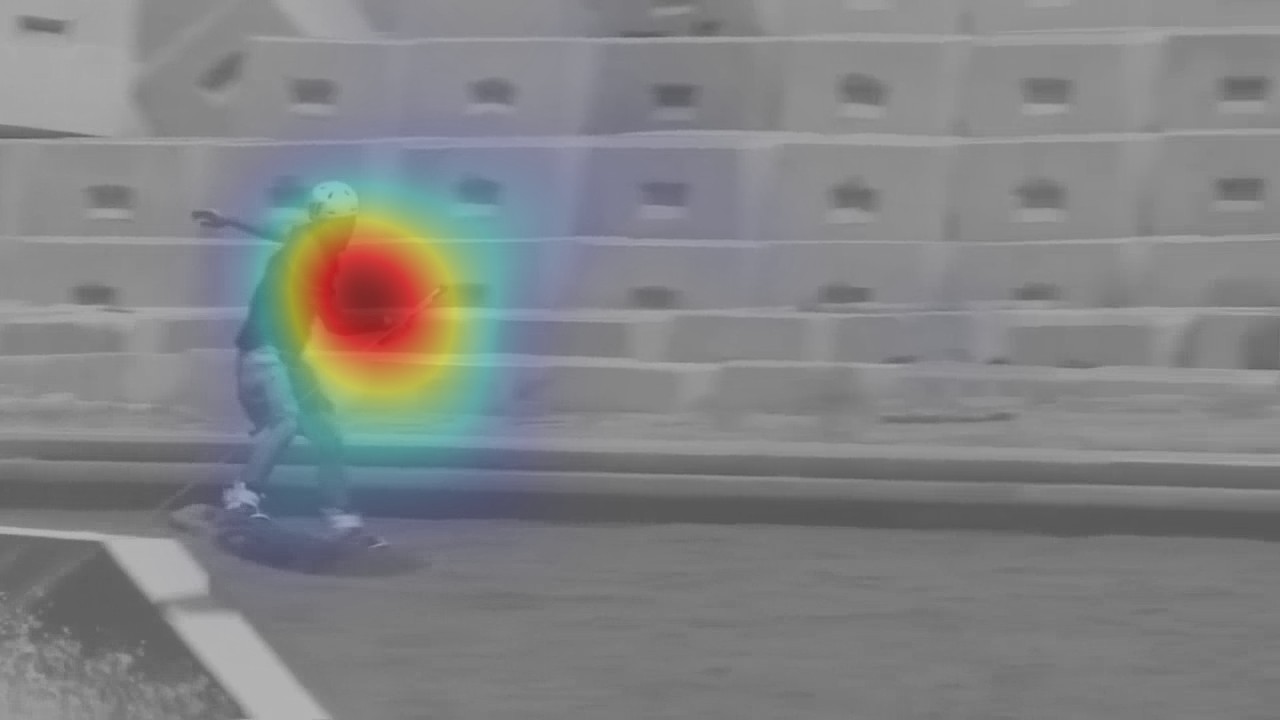} &
\includegraphics[width=5.0cm]{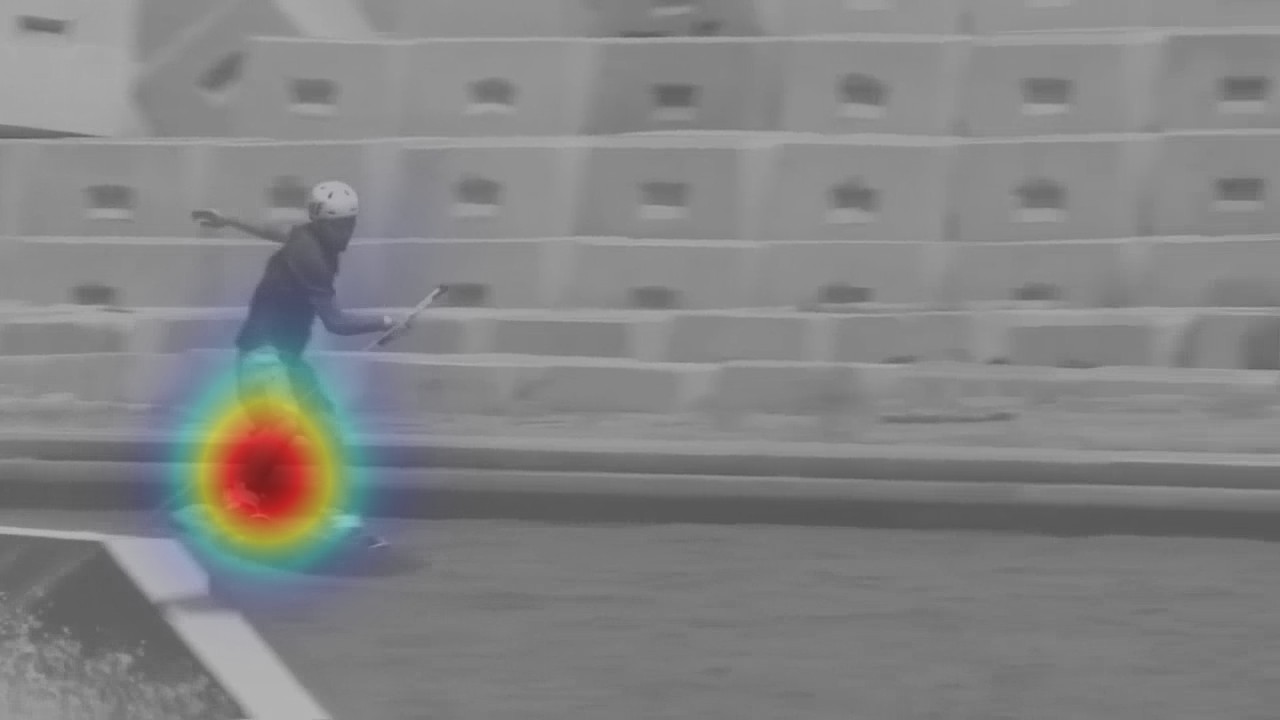} \\
\raisebox{1.4cm}{(11)} &
\includegraphics[width=5.0cm]{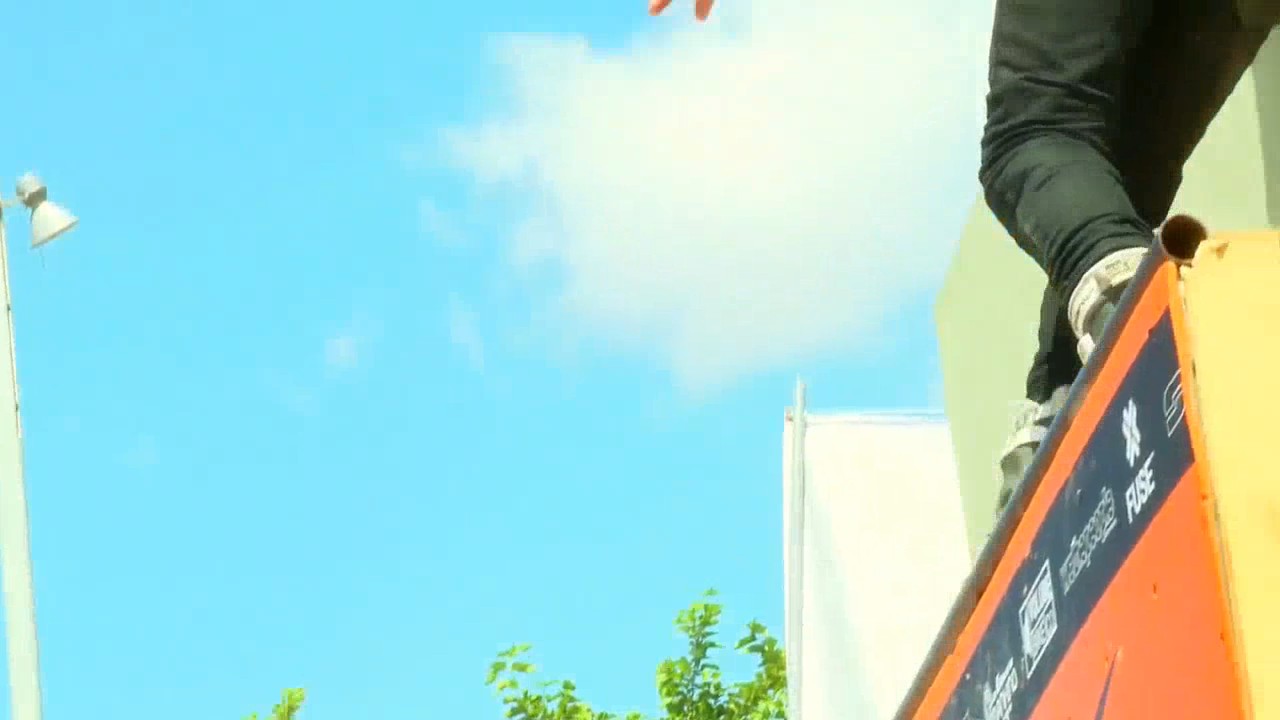} &
\includegraphics[width=5.0cm]{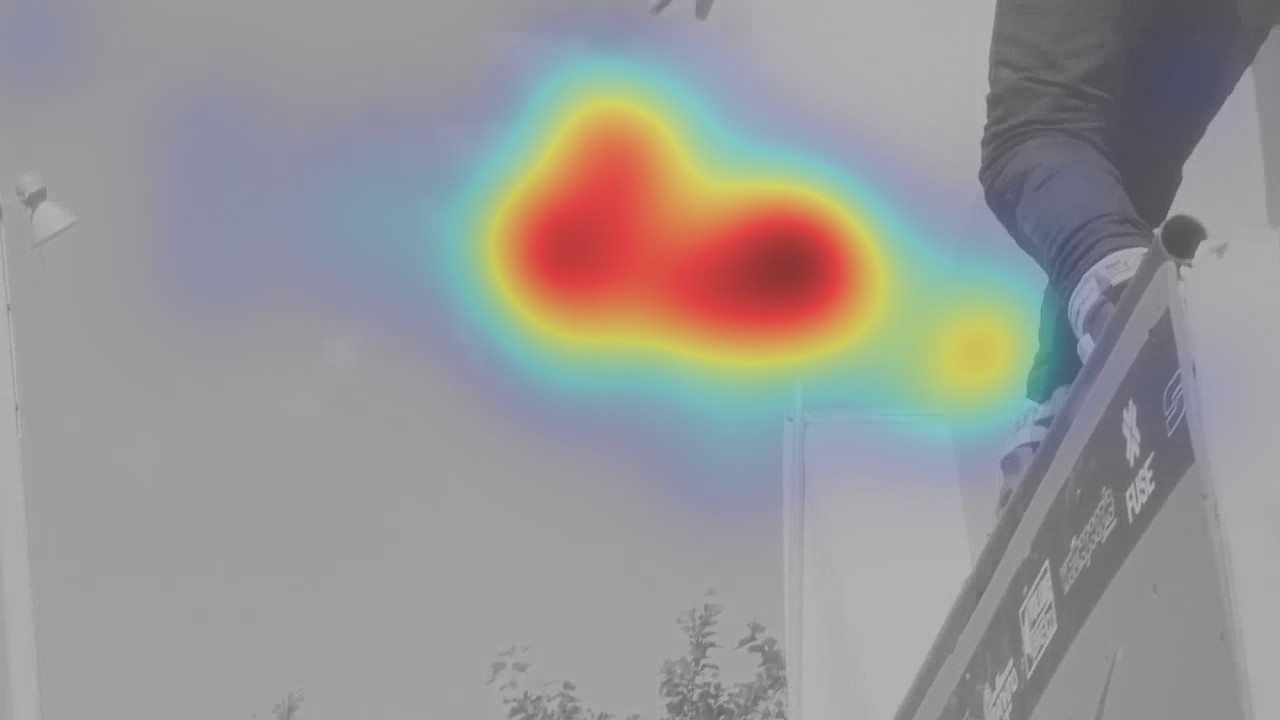} &
\includegraphics[width=5.0cm]{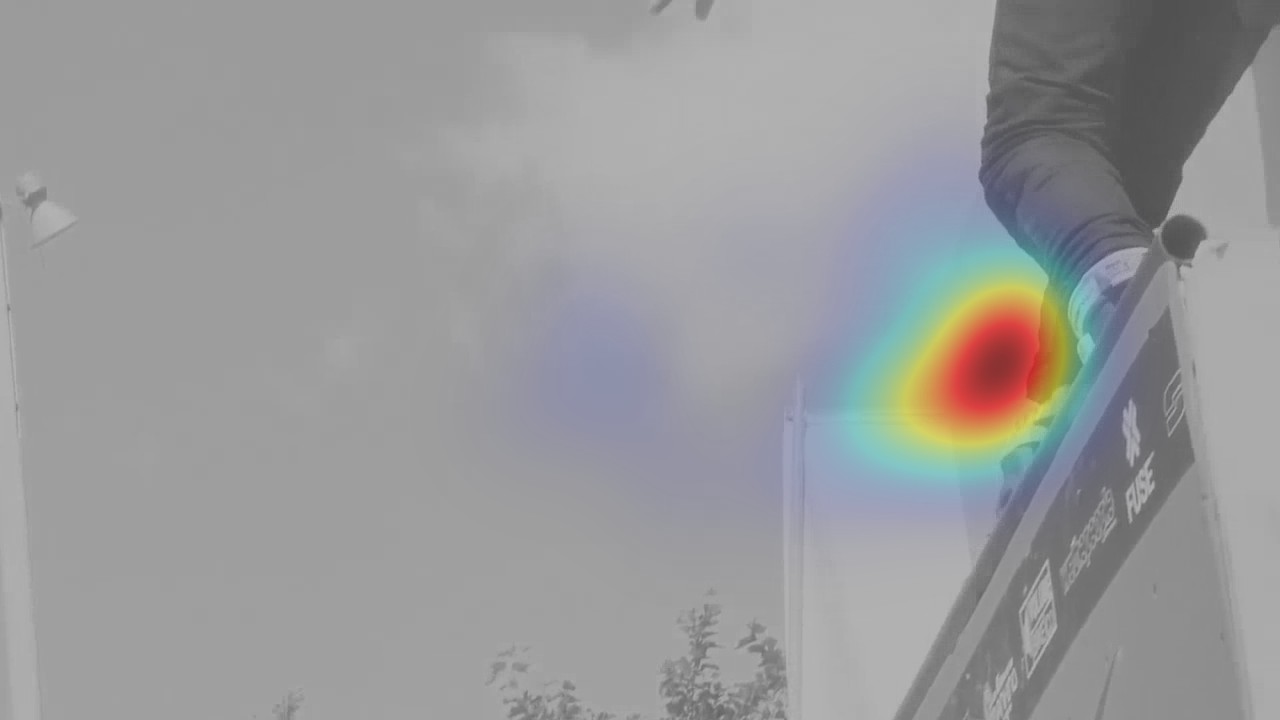} \\
\raisebox{1.4cm}{(12)} &
\includegraphics[width=5.0cm]{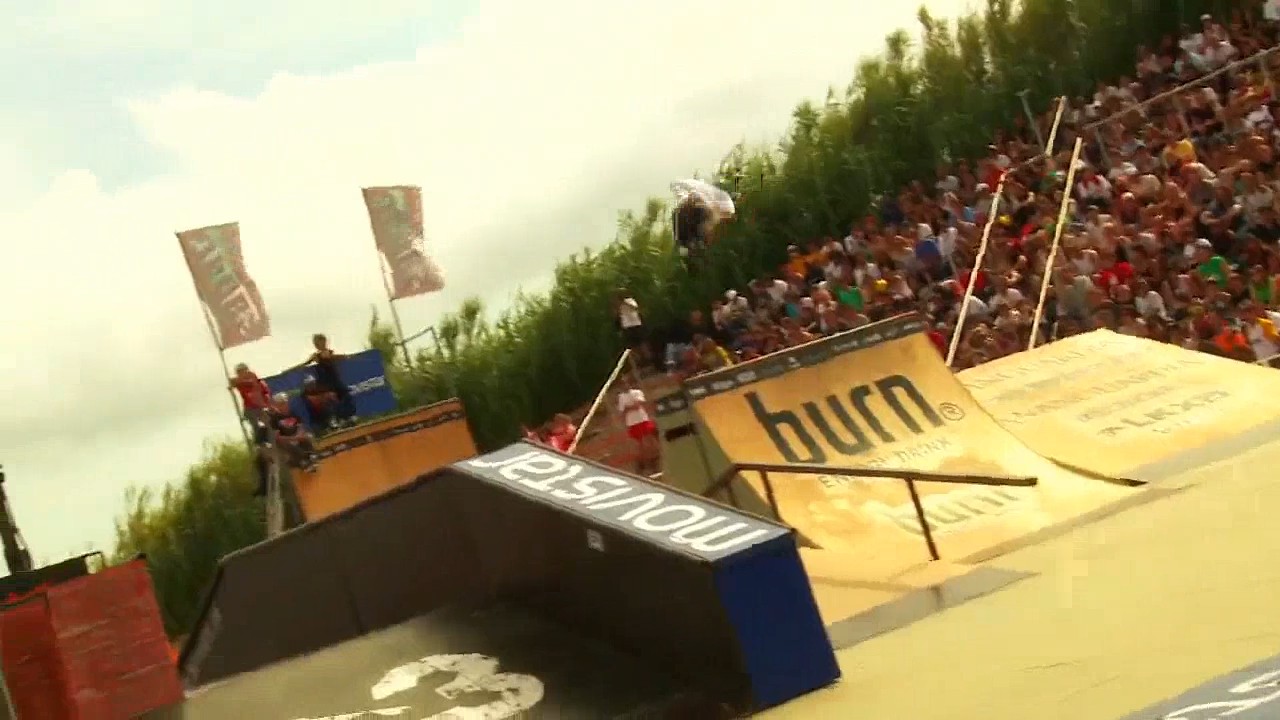} &
\includegraphics[width=5.0cm]{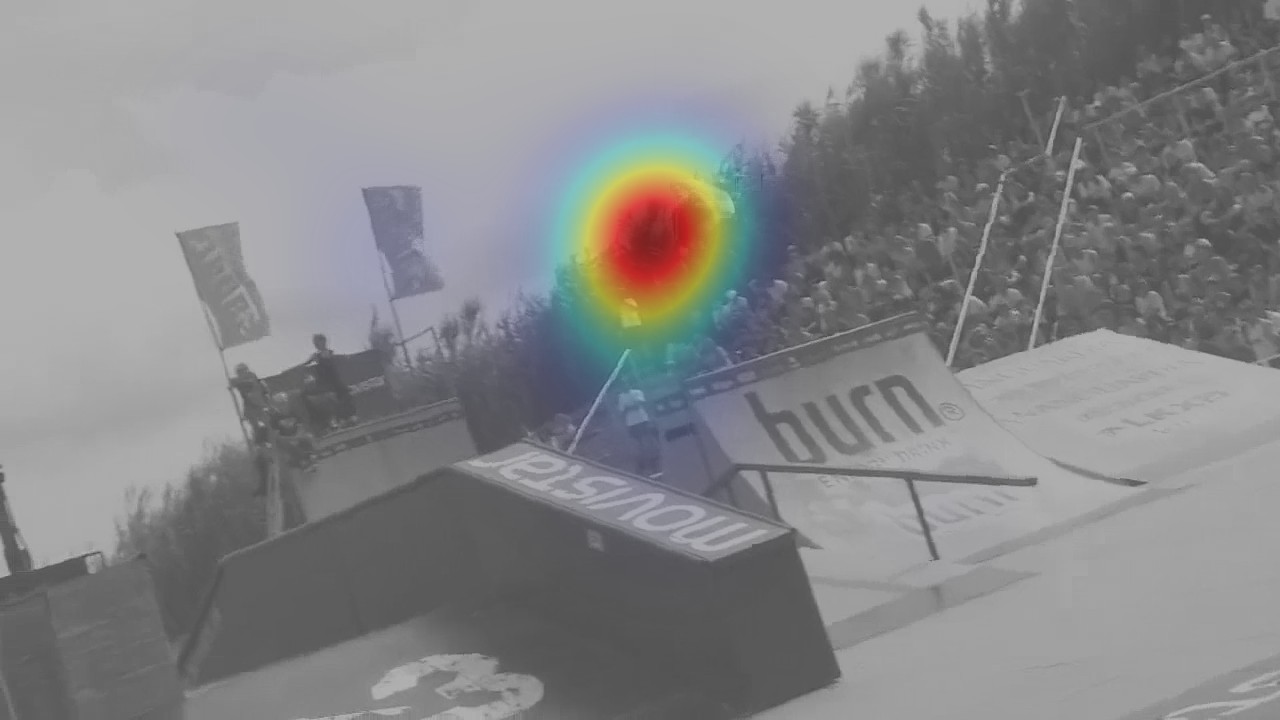} &
\includegraphics[width=5.0cm]{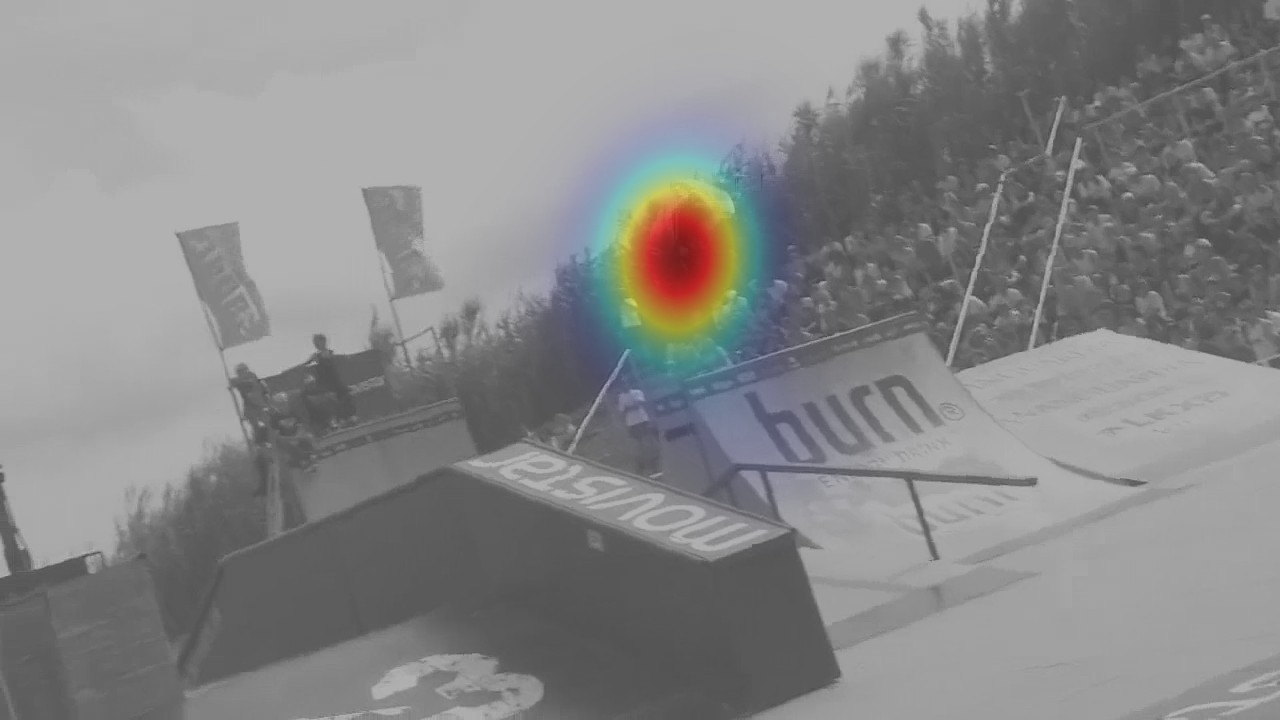} \\
\end{tabular}
  \caption{\label{fig:diem_vis_all_2}
  Additional results for gaze probability estimation. Left to right: original frame, our gaze location data and DIEM's gaze tracking results. The numbers correspond to Figure~\ref{fig:diem_chi-sq}.
}
\end{figure*}

Both our data collection and traditional gaze tracking provide a set of two dimensional points in the frame coordinates. To compare those sets we first estimate a probability density function at every pixel using Kernel Density Estimation (KDE). This is a well-established method for estimating a dense probability function. We use a Gaussian kernel and automatically estimate its variance from the data. Having two probability density functions we compare them using $\chi^2$ distance. The comparison for the frames we used is depicted in Figure~\ref{fig:diem_chi-sq}. The visual comparison for the frames (3) and (5) is shown in Figure~\ref{fig:diem_vis}. All the visualizations can be found in Figure~\ref{fig:diem_vis_all} and Figure~\ref{fig:diem_vis_all_2}.


As one can see, the gaze probability estimated using our data is very close to the gaze tracking data in most of the cases. However, there are specific frames (like 3, 10, 11) where our distribution is relatively far away from the gaze tracking one: all those frames include rapid camera and object motion. One reason for our distribution to be wider in those cases may be that our method's self-reported gaze locations may be offset in time from what a hardware gaze tracker would report, and the gaze direction itself may be altered by
the presentation of the symbol chart.

\subsection{Chart parameters}

The character chart has a number of parameters that can be changed. For some we use constant values in all the experiments. Those include character color and contrast, font size and background color. For the background color we use black, as this is commonly used for letterboxing videos. The characters are $40\%$ gray in all our experiments, to remain readable without introducing excessive contrast. We use $f_s=20px$ font, which is easily readable on most computer and laptop screens.

Another important parameter is character triplet density. Intuitively, we wish them to be as close together as possible to increase the precision of our location measurement. However, placing the triplets too densely will create a very cluttered pattern that is hard to read. To find the optimal triplet density we performed a small experiment, which includes 10 tutorial sessions without video.
We define the {\em relative triplet density} $D_r$ as
\begin{equation}
D_r = \frac{f_s}{d_v}
\end{equation}
where $d_v$ is the vertical distance between the triplets. The horizontal distance is always twice the vertical one $d_h = 2d_v$.
We vary the relative density from $D_r=0.3$ to $D_r=1$ evaluate for the success rate in the tutorials.

\begin{figure}[htb]
    \centering
    \includegraphics[width=0.9\linewidth]{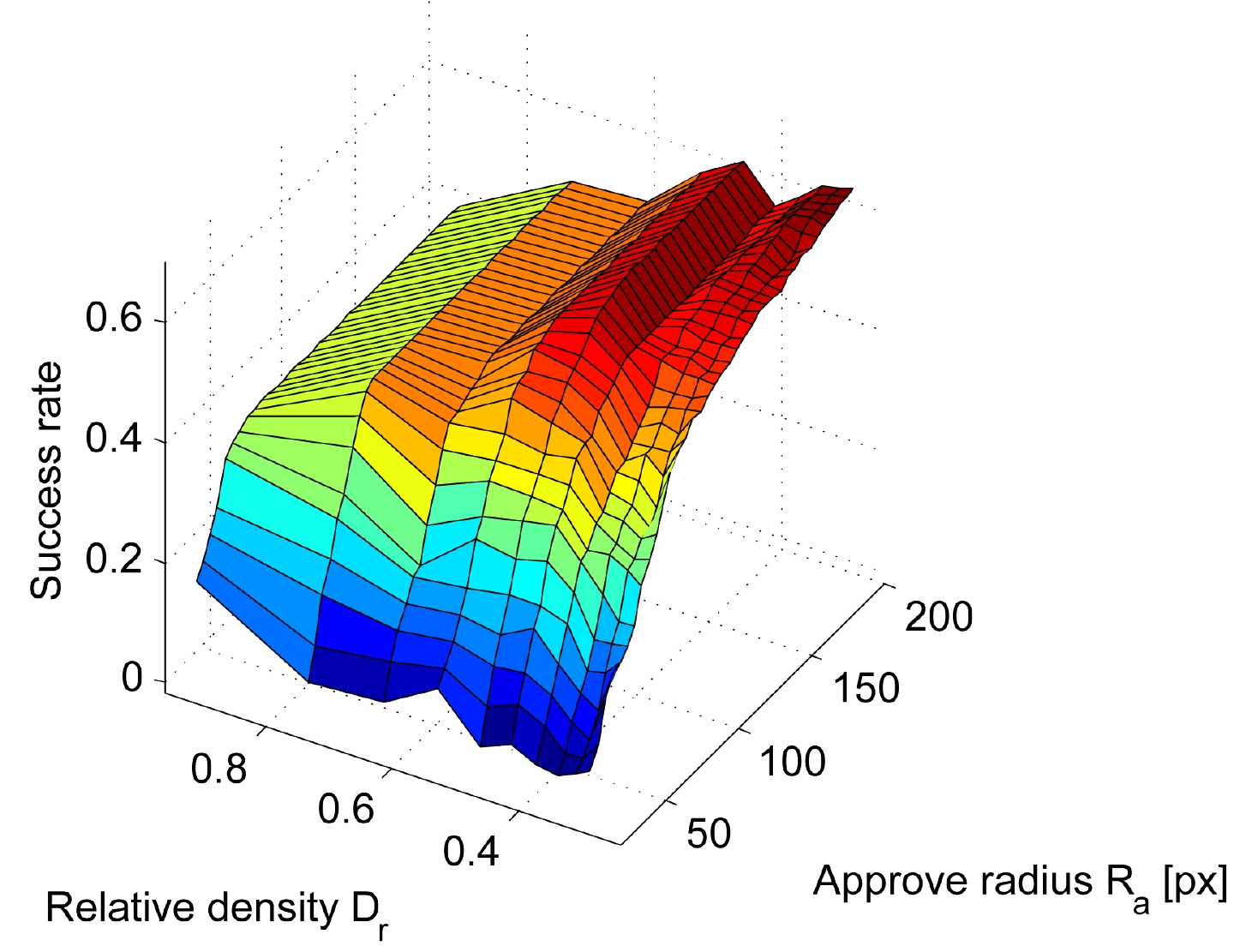}
  \caption{\label{fig:density}
Results for the triplet density experiment. We chose a relative density of 0.5 and approval radius 100 as giving a good tradeoff between precision and recall. The vertical axis shows the success rate of the tutorials.
}
\end{figure}

The results of the triplet density experiments are shown in Figure~\ref{fig:density}. We obtained 1500 samples in this experiment, about 150 per density value.
To assess the success rate of these data, we also vary the approval radius $R_a$ from $20px$ to $200px$.
As one can see, increasing the density beyond $0.5$ leads to a significant decrease in the success rate of the users. As we want the chart to be as dense as possible we set density to $0.5$ for the rest of the experiments described in the paper. This means that the mean distance between triplets is twice the used font size, or $40$ pixels. For the approval radius we see that increasing it beyond some value, that depends on the density, does not improve the results. We set this value to 100 pixels and keep it constant.

\begin{figure}[tbh]
    \centering
    \includegraphics[width=0.9\linewidth]{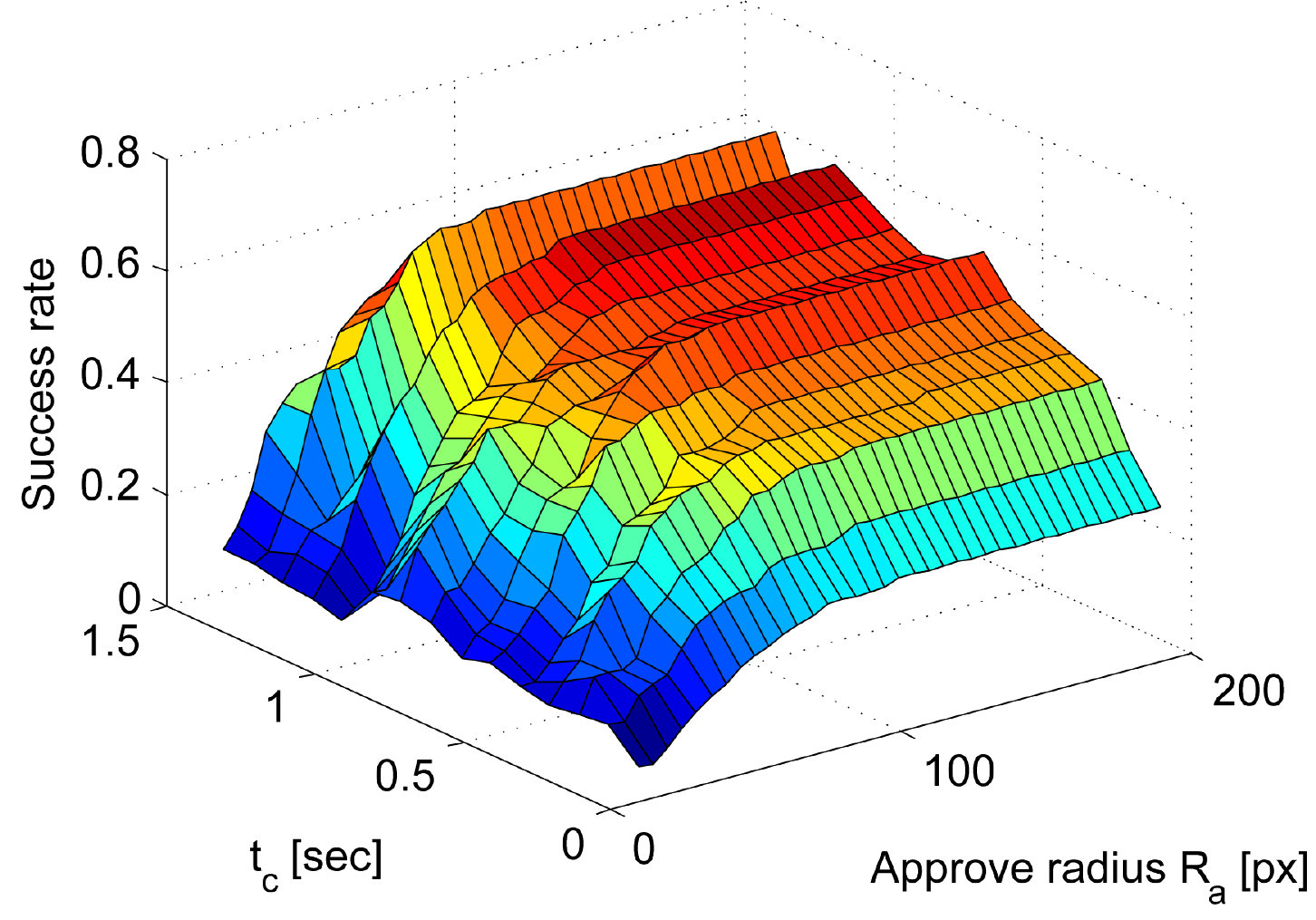}
  \caption{\label{fig:reaction}
Success rate as function of $t_c$. One can see that the optimal duration of chart's stay is about $1sec$. Increasing the approval radius beyond 100 pixels does not increase the performance for all the durations.
}
\end{figure}

An additional parameter we explore is how long the chart should stay on the screen ($t_c$). We performed the same experiment as in the triplet density experiment, but vary the $t_c$ this time. We check the values from $0.1sec$ to $1.5sec$ with steps of $0.1sec$. The results are depicted in Figure~\ref{fig:reaction}. As in the previous experiment, we had about 1500 tutorials, about 100 per $t_c$ value. Here we also vary the approval radius $R_a$ to validate the optimum discovered previously.

As one can see from Figure~\ref{fig:reaction}, the optimal value for $t_c$ is less clear. For very short times we see a clear performance decrease -- it is difficult for humans to read the triplets very fast. However, for long times there is no appreciable improvement in success rate.
One possible reason could be that random saccades start to occur.
Thus, to keep the experiment short and avoid boredom, we choose a value of $t_c=1sec$ and use it in the following experiments. Additionally, the approval radius of $R_a=100px$ appears suitable for any chart duration.

We summarize all the parameters used in our experiments in Table~\ref{tbl:params}
\begin{table}
	\centering
	\caption{\label{tbl:params}
Summary of the parameters used in our experiments.}
	\begin{tabular}{|lcc|}
		\hline
Meaning & Symbol & Value \\ \hline
Clip duration & $t_v$ & $10 sec$ \\
Tutorial duration & $t_t$ & $3 sec$ \\
Chart duration & $t_c$ & $ 1 sec$ \\
Font size & $f_s$ & $20px$ \\
Triplet density & $D_r$ & $0.5$ \\
Approval radius & $R_a$ & $100 px$ \\
		\hline
	\end{tabular}
\end{table}
%


\section{Conclusions}
\label{concl}
In this work we proposed a novel method for capturing the location of the viewer gaze on the screen.
Our system makes it possible to gather gaze locations from an arbitrary number of participants relatively cheaply, but at sparse temporal points. We validated the performance of the proposed method versus traditional gaze tracking methods and found the results comparable. As a drawback, our system does not offer control over the viewing conditions, like glare, distractions or viewing distance. However, depending on the use case, there may be an advantage having gaze locations acquired in random but realistic viewing environments rather than in a perfectly controlled lab setting.

Although it is surprising that self-reported gaze locations can be accurate, we have demonstrated that with the right experimental methodology, self-reported results are statistically similar to those obtained through hardware gaze tracking.  Our contributions lie in the development of this methodology, specifically:
\begin{itemize}
\item using a chart of symbol triplets for explicit location identification and error detection,
\item displaying the symbol chart long enough for legibility but briefly enough to limit the range of eye motion, and
\item using tutorials with an approval radius for screening out lazy participants.
\end{itemize}

We feel that the use of quick tutorials for screening participants is a key contribution of our method.  We observed that even the participants who only passed two tutorials out of ten produced high quality results in the actual experiments.  This suggests that the tutorials encourage the viewers to pay closer attention to their gaze locations.  For this reason, we suspect they are more effective than a qualification test that could be passed just once, but this remains an area of future research.

Our methodology does not control for viewing conditions, because each participant uses his own screen. Hence, our experimental conditions include many uncontrolled variables, including screen contrast, color, resolution, viewing distance, and ambient illumination. This explains why, in some cases, our results differ substantially from those of an in-lab gaze tracking experiment with fixed viewing conditions. For example, watching a video of a dark scene on a screen with low contrast would probably result in attending mostly to the brightest areas, regardless of their content. On the other hand, watching the same video on a bright high-contrast screen could result in attention being drawn to the actions in the video.  Informally we have observed significant changes in gaze when viewing some of our videos at different angles of view.

The flexibility in viewing conditions has both pros and cons. On the down side, it implies that our system is less consistent than a system with controlled viewing conditions. On the up side, our approach better models the reality of personalized viewing:  The distribution of participants' viewing conditions may be more representative of real world video-viewing conditions than a controlled lab experiment.

There is some room for improvement in our experimental design. For example, we employ a chart of  character triplets to capture the user's gaze location, but the current structure of the chart does not cover the entire video uniformly. Specifically, the triplets' location is biased towards the nodes of the regular grid. In future we plan to improve the chart by covering the entire video frame using a Poisson-disk or other less biased distribution.

Although our system is low cost for obtaining sparse samples across many different videos and participants, it would be expensive to employ it for detailed
analysis of a single video on a frame-by-frame basis - this remains better suited to traditional gaze tracking hardware.
However, we believe there are many instances where the advantages of our system outweigh the sparsity of samples. For example, our system can be
deployed across many different geographies.
This allows us to gather gaze location data from all over the world and correlate it to the country of origin of the participants. Note that such experiments would be impossible with traditional gaze tracking approaches. Knowing the relationship between the focus-of-attention and the physical location of the participant will allow more interesting gaze location data analysis.

\section{Acknowledgments}
We would like to thank the DIEM database for making the gaze tracking results publicly available. We additionally thank Cissy Liu for helping you with Flash video advice.
The research of Lihi Zelnik-Manor is supported in part by the Ollendorf foundation, the Israel Ministry of Science, and by the Israel Science Foundation under Grant 1179/11.

\bibliography{dmitry}
%

\end{document}